\documentclass[a4paper,fleqn,usenatbib]{mnras} 

\usepackage[T1]{fontenc}
\usepackage{ae,aecompl}

\usepackage{geometry}
\usepackage{nicefrac}
\usepackage{float}    
\usepackage{subfig}   
\usepackage{subfloat} 

\usepackage{graphicx}	
\usepackage{amsmath}	
\usepackage{amssymb}	

 \title{Calibration of Herschel SPIRE FTS observations at different spectral resolutions}

 \author[N.~Marchili et al.]{N.~Marchili$^{1}$\thanks{E-mail: nicola.marchili@gmail.com}, R.~Hopwood$^{2}$, T.~Fulton$^{3, 4}$, E.~T.~Polehampton$^{3, 5}$, I.~Valtchanov$^{6}$,  \newauthor
 J.~Zaretski$^{4}$, D.~A.~Naylor$^{3}$, M.~J.~Griffin$^{7}$, P.~Imhof$^{3, 4}$, T.~Lim$^{5, 6}$, N.~Lu$^{8, 9}$,\newauthor
 G. Makiwa$^{3}$, C.~Pearson$^{5, 10, 11}$, L.~Spencer$^{3}$
\\
$^{1}$IAPS-INAF, Via Fosso del Cavaliere 100, 00133, Roma, Italy\\
$^{2}$Department of Physics, Imperial College London, Prince Consort Road, London SW7 2AZ, UK\\
$^{3}$Institute for Space Imaging Science, Department of Physics \& Astronomy, University of Lethbridge, 4401 University Drive, Lethbridge, Alberta, T1K 3M4, Canada\\
$^{4}$Blue Sky Spectroscopy, Lethbridge, AB, T1J 0N9, Canada\\
$^{5}$RAL Space, Rutherford Appleton Laboratory, Chilton, Didcot, Oxfordshire, OX11 0QX, UK\\
$^{6}$European Space Astronomy Centre, Herschel Science Centre, ESA, 28691 Villanueva de la Ca\~nada, Spain\\
$^{7}$School of Physics and Astronomy, Cardiff University, The Parade, Cardiff, CF24 3AA, UK\\
$^{8}$China-Chile Joint Center for Astronomy, Chinese Academy of Sciences, Camino El Observatorio, 1515 Las Condes, Santiago, Chile\\
$^{9}$National Astronomical Observatories of China, Chinese Academy of Sciences, Beijing 100012, China\\
$^{10}$Department of Physical Sciences, The Open University, Milton Keynes, MK7 6AA, UK\\
$^{11}$Oxford Astrophysics, Denys Wilkinson Building, University of Oxford, Keble Rd, Oxford OX1 3RH, UK
}

\date{Accepted XXX. Received YYY; in original form ZZZ}

\pubyear{2016}

\begin{document}
\label{firstpage}
\pagerange{\pageref{firstpage}--\pageref{lastpage}}
\maketitle

\begin{abstract}
The SPIRE Fourier Transform Spectrometer on board the Herschel Space Observatory had
two standard spectral resolution modes for science observations: high resolution (HR) and low resolution (LR),
which could also be performed in sequence (H+LR). A comparison of the HR and LR resolution
spectra taken in this sequential mode, revealed a systematic discrepancy in the continuum level. Analysing the data at different stages during standard pipeline processing, demonstrates the telescope and instrument emission affect HR and H+LR observations in a systematically different way. The origin of this difference is found to lie in the variation of both the telescope and instrument response functions, while it is triggered by fast variation of the instrument temperatures. As it is not possible to trace the evolution of the response functions through auxiliary housekeeping parameters, the calibration cannot be corrected analytically. Therefore an empirical correction for LR spectra has been developed, which removes the systematic noise introduced by the variation of the response functions.
\end{abstract}




\section{Introduction}

The Spectral and Photometric REceiver (SPIRE; \citealt{Griffin10}) was one of three focal plane instruments on board the ESA Herschel Space Observatory (Herschel; \citealt{Pilbratt10}). The instrument consisted of an imaging photometric camera and an imaging Fourier Transform Spectrometer (FTS). Both sub-instruments used bolometric detectors operating at $\sim300$ mK (\citealt{Turner01}) with feedhorn focal plane optics giving sparse spatial sampling over an extended field of view (\citealt{Dohlen00}). The FTS had two broad-band intensity beam splitters in a Mach-Zehnder configuration (\citealt{Ade99}; \citealt{Swinyard03}) and two bolometer arrays with partially overlapping bands: the SPIRE Long Wavelength spectrometer array (SLW; 447-1018 GHz) and SPIRE Short Wavelength spectrometer array (SSW; 944-1568 GHz), with 19 and 37 detectors. Detailed information about the standard procedure for the calibration of FTS data and the calibration accuracy can be found in \cite{Swinyard10}, \cite{Swinyard14}, and \cite{Hopwood15}. All products in this paper were produced by the standard data processing pipeline (Fulton et al. 2016) in the Herschel Interactive Processing Environment (HIPE; \citealt{Ott10}) version 13 and calibration tree \textsc{spire\_cal\_13\_2}.

The FTS operated by splitting incoming radiation into two beams. An optical path difference (OPD) was introduced between the beams by scanning an internal mirror, so that when recombined, an interference pattern (known as an interferogram) is formed. The inverse Fourier transform of the interferogram gives the spectrum, with the spectral resolution set by the
maximum OPD between the interfering beams. The FTS observed with two standard spectral resolutions for science observations: high resolution (HR, at 1.184 GHz spectral resolution), low resolution (LR, at 24.98 GHz) and a combination of the two performed in a sequence, called H+LR. There were two additional operating modes: medium resolution (MR, at 7.2 GHz) and calibration resolution (CR). CR had the same spectral resolution as HR, but provided measurements out to greater negative OPD, which in turn allowed the phase to be determined to greater spectral resolution and thus provide improved phase correction. This mode was only used for calibration observations and is processed as HR by the pipeline (see \citealt{Fulton16}). While MR was never used for science observations and only occasionally for calibration. It is calibrated as the LR mode (see \citealt{SPIRE16} for more details).

The FTS had an internal beam steering mirror (BSM) which provided different spatial sampling (\citealt{Fulton10}): sparse (the BSM is fixed at its home position); intermediate (the BSM samples 4 spatial positions equivalent to full beam sampling); full (the BSM samples 16 positions achieving Nyquist sampling). In this paper we only consider sparse mode observations, while the corrections introduced later on apply to all spatial sampling modes.

\begin{figure}
   \centering
   \includegraphics[width=0.96\columnwidth]{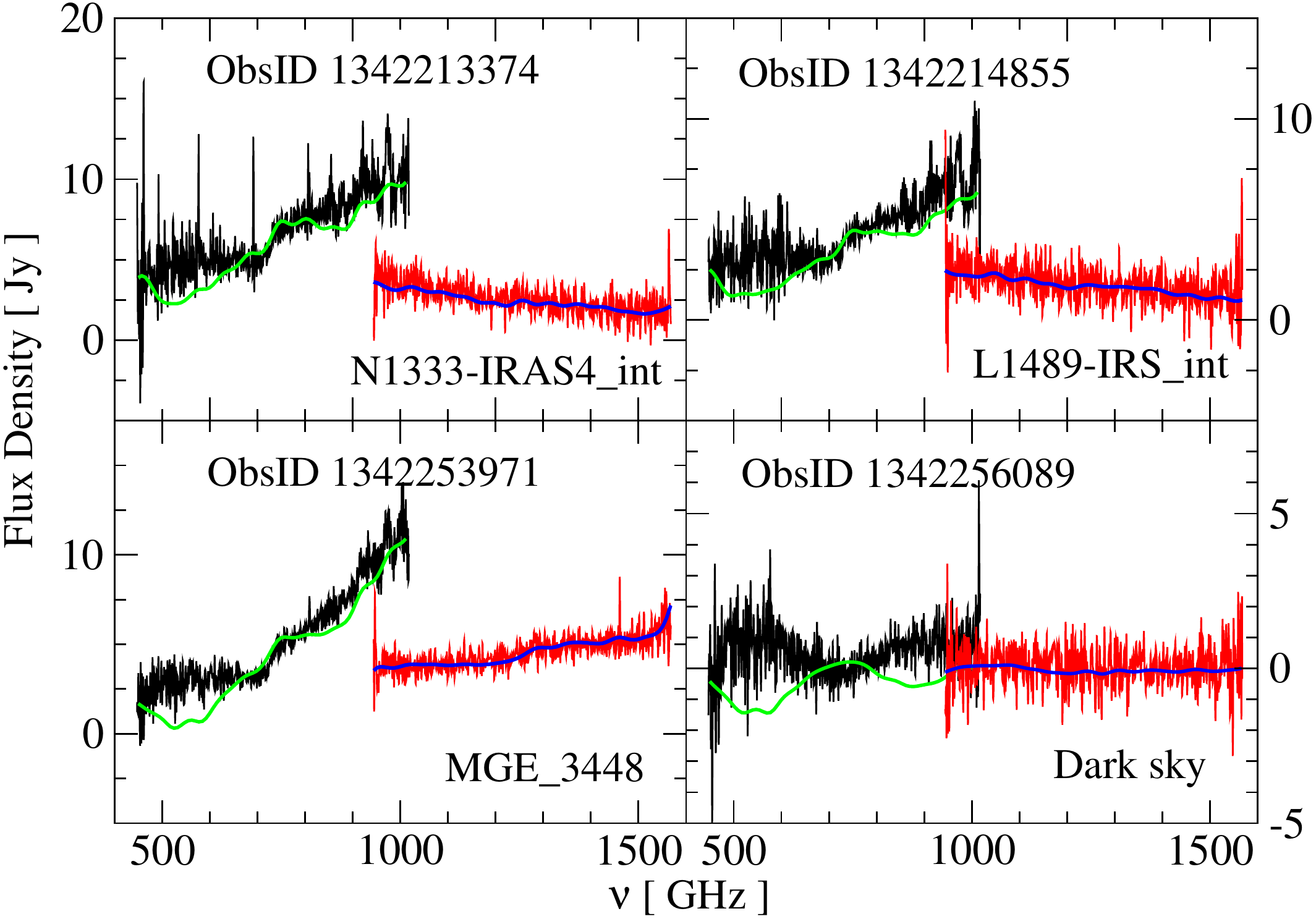}
      \caption{Some examples of H+LR observations. In green and blue, the low-resolution spectra from detectors SLWC3 and SSWD4, respectively; in black and red, the high-resolution ones.}
              \label{fig:ex1}

   \centering
   \includegraphics[width=0.96\columnwidth]{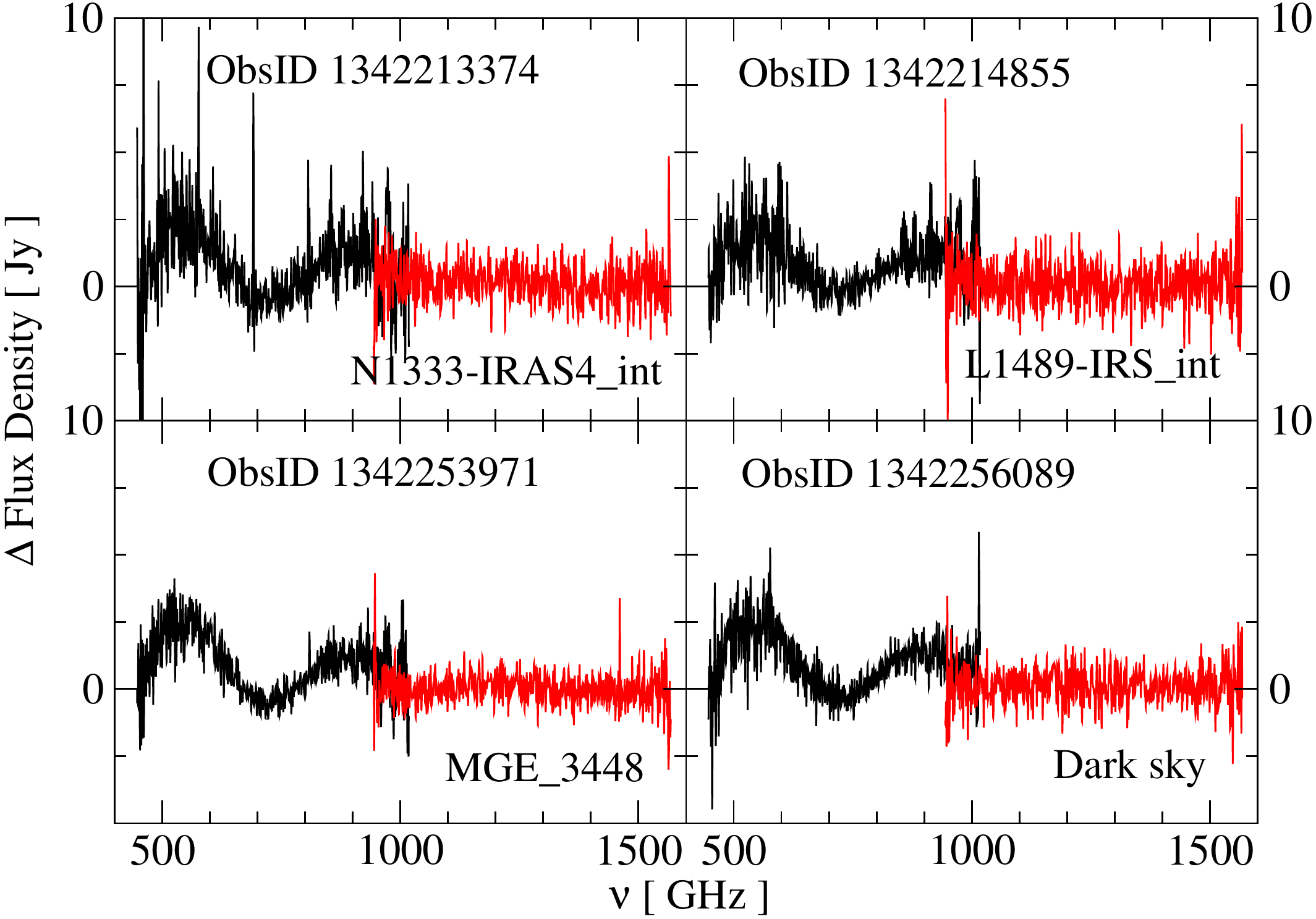}
      \caption{The difference between the high-resolution and the low-resolution spectra shown in Fig. \ref{fig:ex1}.}
              \label{fig:ex2}
\end{figure}

The main goal for observations in H+LR mode was to allow for better sensitivity on the continuum, while minimising the observing time on the more time intensive HR mode. However,  comparing the final point-source calibrated spectra for the HR part and the LR part of H+LR observations (from now on, H+LR(H) and H+LR(L), respectively) reveals significant discrepancies between the respective modes. This discrepancy, which was found also in spectra calibrated with previous versions of HIPE, is independent of the target observed and is more evident in the low frequency detector array (SLW). Fig. \ref{fig:ex1} shows several examples of H+LR(H) and H+LR(L) spectra, with the corresponding difference shown in Fig. \ref{fig:ex2}. These figures illustrate the systematic nature of the discrepancy, which takes the form of a characteristic double bump, peaking around 550 and 900 GHz.


We present an analysis of the discrepancies seen in the spectra as a function of operating mode. Figures \ref{fig:ex1} and \ref{fig:ex2} show that the problem primarily concerns spectra from SLW, so we focus on the centre SLW detector (SLWC3). The basic characteristics of the signal distortion are identified in Sec. 2 and 3. Several hypotheses concerning the nature of the problem are presented in Sec. 4, 5, and 6, where the most plausible explanation is discussed in detail. Since there is no possible analytical solution to the problem, an empirical correction has been developed (Sec. 7). Its effectiveness is discussed in Sec. 8 through the comparison of the calibration uncertainty before and after the correction. The main findings of this work are summarised in Sec. 9.

\section{Comparison of uncalibrated spectra}
\label{Comparison}


The total uncalibrated signal is the sum of three components: the telescope emission, the instrument emission, and the source emission.
The first two contributions can be expressed in terms of voltage density as

\begin{equation}
V_\mathrm{Tel}(\nu)=M_\mathrm{Tel}(\nu) R_\mathrm{Tel}(\nu) $\hspace{0.2cm} [V GHz$^{-1}]
\end{equation}
and

\begin{equation}
V_\mathrm{Inst}(\nu)=M_\mathrm{Inst}(\nu) R_\mathrm{Inst}(\nu) $\hspace{0.2cm} [V GHz$^{-1}]
\end{equation}
where $M_\mathrm{Tel}(\nu)$ and $M_\mathrm{Inst}(\nu)$ are the telescope and instrument models, which can be calculated from housekeeping parameters (\citealt{Fulton16}), and $R_\mathrm{Tel}(\nu)$ and $R_\mathrm{Inst}(\nu)$ are the telescope and instrument relative spectral response functions (RSRFs; see \citealt{Fulton14}). 

$R_\mathrm{Tel}(\nu)$ and $R_\mathrm{Inst}(\nu)$ depend on resolution mode. The estimation of the response functions is performed through a procedure that aims to minimise the residual noise in observations of the SPIRE dark sky field. This is a dark region of sky centred on RA:17h40m12s and Dec:+69d00m00s (J2000) and selected on the grounds that the region has: low cirrus, is visible at all times, and contains no SPIRE-bright sources. Before HIPE 9, LR calibration was extrapolated from the low resolution portion of calibration resolution (CR) and HR dark sky observations. This method, however, was not optimal for LR observations, as large systematic residuals were left in the calibrated spectra. For the later versions of HIPE, the calibration of LR spectra is carried out using a new set of response functions, based on LR dark sky observations. The HR RSRFs, instead, are calculated by considering HR dark sky observations.

The telescope and instrument contributions to the signal are removed at two key steps of the calibration pipeline. The existence of a difference between the response functions for HR and LR data implies that the amount of signal removed at these stages of the data calibration changes with the resolution mode, raising the question as to whether the discrepancies shown in Fig. \ref{fig:ex2} are intrinsic to raw data, or are introduced during processing. 

\begin{figure}
   \centering
   \includegraphics[width=0.96\columnwidth]{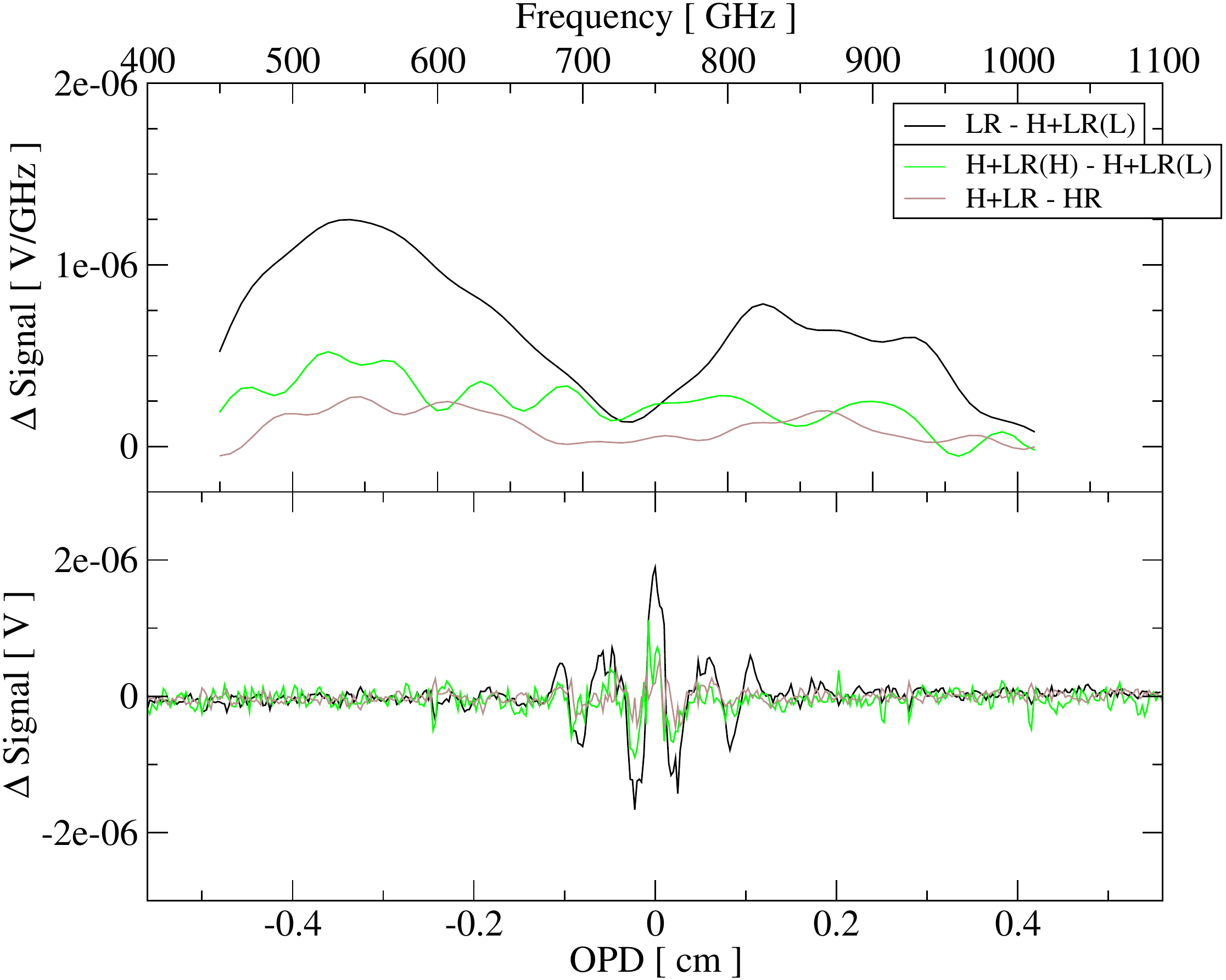}
      \caption{Upper panel: an example of the differences between the uncalibrated spectra obtained at different resolution modes. Lower panel: the differences seen in the interferograms. Uncertainties on the plotted quantities are below 1\%.
                    }
              \label{fig:sdi-fts}
\end{figure}

The answer comes from the comparison between uncalibrated H+LR(H) and H+LR(L) spectra. Their difference before the telescope and instrument correction is marginal. It also appears that HR uncalibrated spectra are consistent with both H+LR(H) and H+LR(L) spectra, while the LR ones are systematically different from all the others (see upper panel of Fig. \ref{fig:sdi-fts}). Therefore, the root problem is not with H+LR spectra, rather it is the systematic difference between LR spectra and the ones obtained at any other resolution mode, from now on indicated as HR/H+LR. Looking at the difference between LR and H+LR(L) (black line in Fig. \ref{fig:sdi-fts}), the double bump already evident in Fig. \ref{fig:ex2} is clearly discernible. 
Significant discrepancies between LR and H+LR data concern not only the spectra, but also the interferograms (see lower panel of Fig. \ref{fig:sdi-fts}), and Spectrometer Detector Timelines (SDTs), suggesting a difference that must exist in the raw signal.

It can be inferred that, while the $R_\mathrm{Inst}(\nu)$ function implemented in the pipeline is efficient in the calibration of LR spectra, it introduces a systematic bias in the calibration of the H+LR(L) spectra. It should be noted, however, that such bias does not exclusively affect H+LR(L), as its characteristic shape can also be recognised in LR spectra, albeit with lower amplitudes.

\section{Analysis of dark sky observations}

A fundamental issue must be addressed when investigating the origin of the discrepancy between HR/H+LR and LR spectra: is the difference between these spectra a constant or does it depend on some parameters, such as time, instrument temperature, or telescope temperature? The differences calculated from several pairs of observations (as it will be shown in Sec. \ref{Dark}) are not identical to each other. The observed variations can be due to random noise, or to other causes that have nothing to do with the resolution modes in which the observations were taken (see \citealt{Swinyard14}). Besides these contributions, there are variations specifically relating to the resolution mode of the data, and the root cause of this discrepancy has been investigated.

In the following, we will calculate the differences between spectra for specific pairs of observations; these differences will be referred to as $\delta_\mathrm{HR-LR}(\mathrm{ObsID}_\mathrm{HR/H+LR}, \mathrm{ObsID}_\mathrm{LR})$, where ObsID stands for the identification numbers of the observations. The systematic difference between HR/H+LR and LR spectra, defined as the {\emph{ideal} difference between spectra observed simultaneously and free from noise and resolution-independent sources of radiation, will be referred to as $\delta_\mathrm{HR-LR}$.

The difference $\delta_\mathrm{HR-LR}$ was estimated for a set of dark sky observations performed at different resolution modes. Given the low number of H+LR observations and the similarity between H+LR and HR data, we focused on the comparison between HR and LR dark sky observations. Note that HR observations can be calibrated as if they were taken in LR mode, by truncating HR interferograms at the same OPD for LR. While true in principle, this may not be valid in practice, as the heat dissipated during the high-resolution scan (which goes as OPD$^2$) is much greater and results in a different instrument environment. Most of the LR dark sky observations acquired by the SPIRE FTS are concentrated within the time span between the Operational Days (OD) 1079 and 1433. Before OD 1079, dark sky observations were generally taken in CR mode, which is analogous to the HR mode (see the \citealt{SPIRE16}). For the present analysis, we focused on a sample of dark sky observations between OD 1079 and OD 1325. The identification numbers of these observations are listed in Table \ref{tab:obsid}.

\begin{table}
\caption{Dark sky observations used for the comparison of spectra at different resolution modes. Col. 1 reports the operational day. Columns 2 and 3 give the identification number of the LR and the HR observations, respectively.\label{tab:obsid}}  
\centering
\begin{tabular}{c c c}
\hline
\noalign{\smallskip}
 OD & ObsID$_\mathrm{LR}$ & ObsID$_\mathrm{HR}$ \\
\hline
\noalign{\smallskip}
1079 &  1342245124 & 1342245125\\
1098 &  1342245852 & 1342245853\\
1111 &  1342246260 & 1342246261\\
1125 &  1342246983 & 1342246984\\
1130 &  1342247108 & 1342247109\\
1144 &  1342247574 & 1342247575\\
1150 &  1342247752 & 1342247753\\
1160 &  1342248234 & 1342248235\\
1177 &  1342249067 & 1342249068\\
1186 &  1342249453 & 1342249454\\
1207 &  1342250517 & 1342250518\\
1262 &  1342253973 & 1342253972\\
1283 &  1342255268 & 1342255269\\
1291 &  1342256085 & 1342256091\\
1291 &  1342256086 & 1342256091\\
1291 &  1342256088 & 1342256091\\
1298 &  1342256360 & 1342256361\\
1313 &  1342257334 & 1342257335\\
1325 &  1342257920 & 1342257921\\
\hline
\end{tabular}
\end{table}

From the comparison of the uncalibrated spectra (see Fig. \ref{fig:lr-hr}), it is easy to recognise the double bump that characterises the discrepancy between HR and LR data. However, it is also evident that there is a large spread in the plotted curves, caused, as previously mentioned, by the significant differences in the instrument emission between the LR and the quasi-simultaneous HR observation \footnote{The important contribution of the instrument emission to the difference between LR and HR spectra might appear in contradiction with the negligible difference seen between H+LR(H) and H+LR(L), as reported in the previous section. The two cases, however, are different, because of the order in which the observations are taken. LR observations are almost always taken before HR ones, while, in H+LR observations, the low-resolution scans always follow the high-resolution ones. This point will be further considered later on.}. On the one hand, in order to isolate the systematic part of the discrepancy from possible contamination introduced by the instrument and telescope corrections, it would seem reasonable to look at uncalibrated spectra. On the other hand, the differences in the instrument emission between consecutive observations are such that, without instrument correction, $\delta_\mathrm{HR-LR}$ can not be properly evaluated. The similarity between the spectra obtained after the instrument correction (see Fig. \ref{fig:lr-hr2}) suggests that this is the optimal stage of calibration to estimate $\delta_\mathrm{HR-LR}$.

The modest fluctuations among the $\delta_\mathrm{HR-LR}$ curves obtained at different ODs indicates that, to a first approximation, the discrepancy between LR and HR/H+LR spectra can be regarded as constant in time. In the standard calibration procedure (see Sec. \ref{Comparison}), the total signal is regarded as the sum of three components, one depending on the target of the observation, the second on telescope temperature, the third one on instrument temperature. Since the three components are variable, it is hard to understand how any of these can contribute a spurious, resolution dependent signal that is approximately constant in time. To check whether the problem can be resolved with an empirical modification to the calibration, we consider a third noise component that only depends on resolution modes and is added to the noise contributed by the telescope and the instrument.


\begin{figure}
   \centering
   \includegraphics[width=0.96\columnwidth]{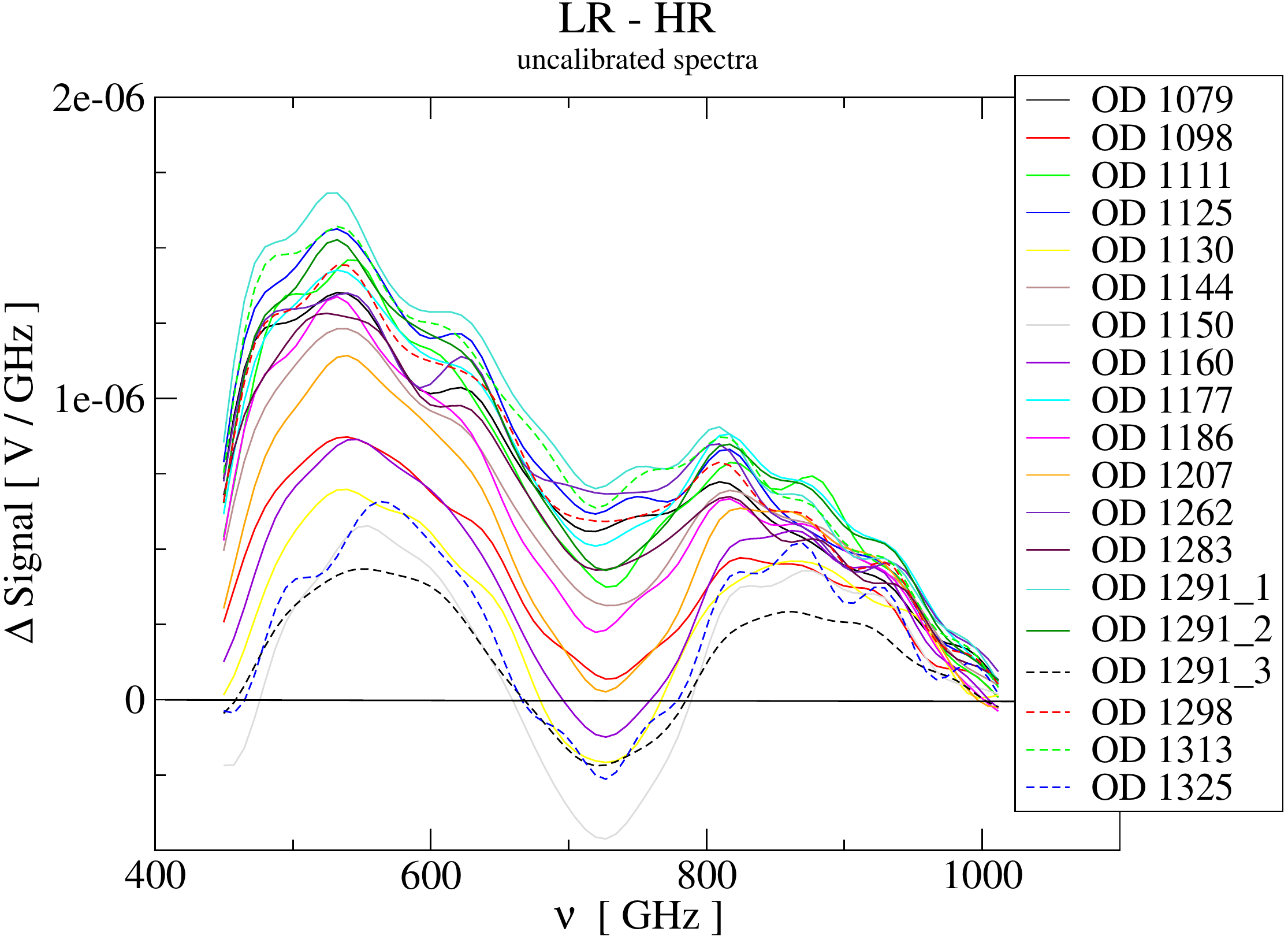}
      \caption{Difference between quasi-simultaneous LR and HR spectra. The relatively large spread between the lines is caused by the uncorrected instrument emission. Uncertainties on the plotted quantities are below 1\%.}
              \label{fig:lr-hr}

   \centering
   \includegraphics[width=0.96\columnwidth]{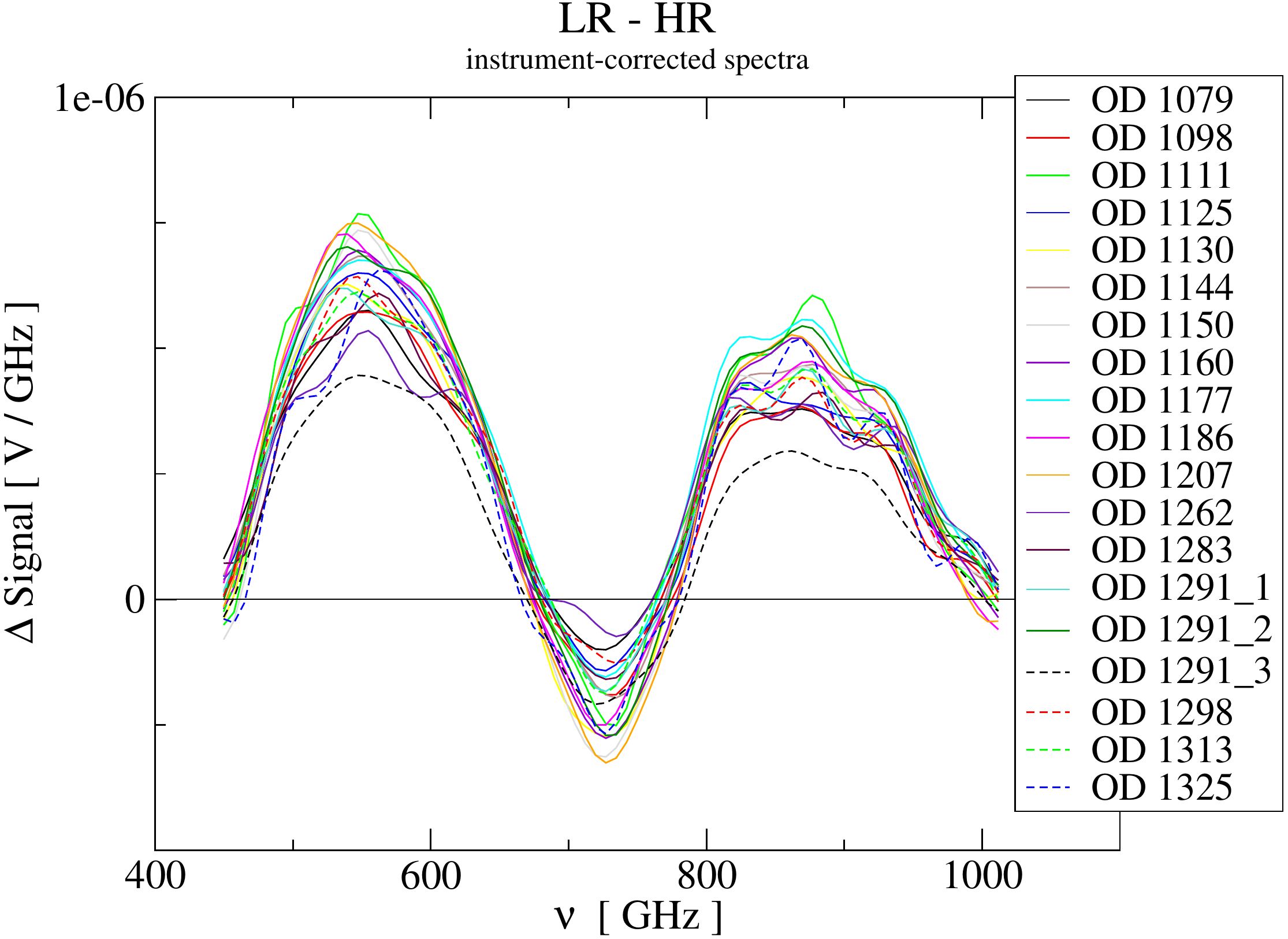}
      \caption{Difference between quasi-simultaneous LR and HR spectra, after applying the instrument correction discussed in Sec. \ref{Comparison}. Uncertainties on the plotted quantities are below 1\%.}
              \label{fig:lr-hr2}
\end{figure}

\subsection{Three-parameter model for the calibration of the data}
\label{Three-parameter}


The three-parameter calibration model assumes that the measured voltage density $V_i$ for a dark sky observation $i$ can be expressed as

\begin{equation} \label{eq:three-par}
V_i(\nu)=M_\mathrm{Tel}(\nu) {R'_\mathrm{Tel}}(\nu)+M_\mathrm{Inst}(\nu) {R'_\mathrm{Inst}}(\nu)+f(\nu).
\end{equation}
where ${R'_\mathrm{Tel}}(\nu)$ and ${R'_\mathrm{Inst}}(\nu)$ are new estimates of the telescope and instrument response functions, and $f(\nu)$ is a resolution dependent calibration parameter.
For each of the observations in Table \ref{tab:obsid}, the variables $V_i(\nu)$, $M_\mathrm{Tel}(\nu)$, and $M_\mathrm{Inst}(\nu)$ can be determined. Given the 19 LR (17 HR) observations at our disposal, for each frequency a system of 19 (17) equations with three unknowns can be build to calculate the best-fit parameters for LR (HR) data. Since the systems are overdetermined, a least-square fitting algorithm was used to simultaneously estimate ${R'_\mathrm{Tel}}(\nu)$, ${R'_\mathrm{Inst}}(\nu)$, and $f(\nu)$. The results are plotted in Fig. \ref{fig:model}.

\begin{figure}
   \centering
   \includegraphics[width=0.96\columnwidth]{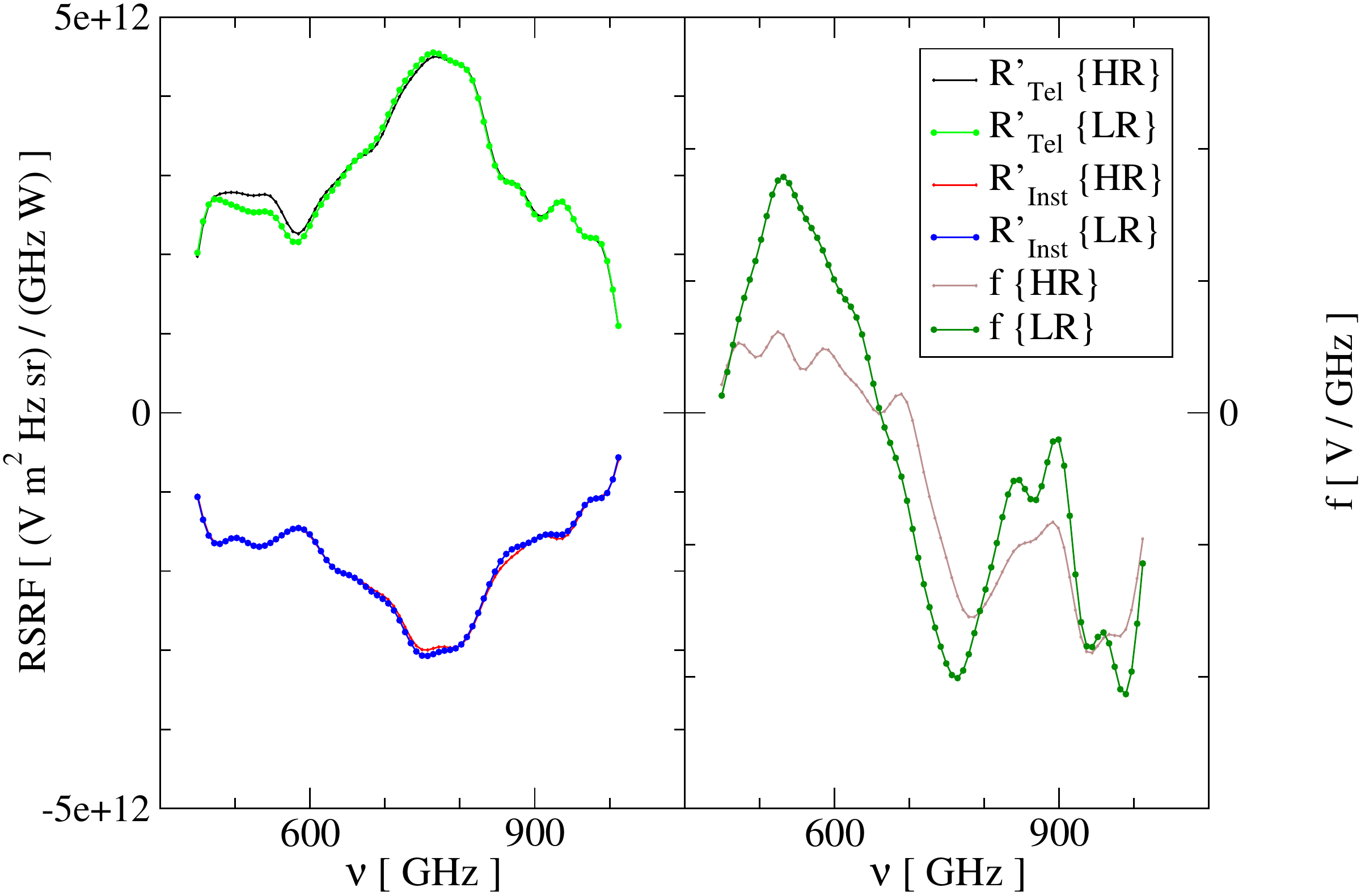}
      \caption{The best fit RSRFs (left panel) and the $f(\nu)$ parameter (right panel) estimated from the 19 LR and 17 HR dark sky observations in Table \ref{tab:obsid}. The HR spectra utilised for calculating the parameters have been produced in low-resolution mode using the specific option in the data reduction pipeline.}
              \label{fig:model}
\end{figure}

The differences between the best fit parameters for LR and HR data are not limited to $f(\nu)$. Similarly to the standard pipeline, we find that the RSRFs at different resolutions are inconsistent. In Fig. \ref{fig:cor-dif}, the differences between the contributions at different resolution modes provided by each of the three components on the right-hand side of Eq. \ref{eq:three-par} are shown.

\begin{figure}
   \centering
   \includegraphics[width=0.96\columnwidth]{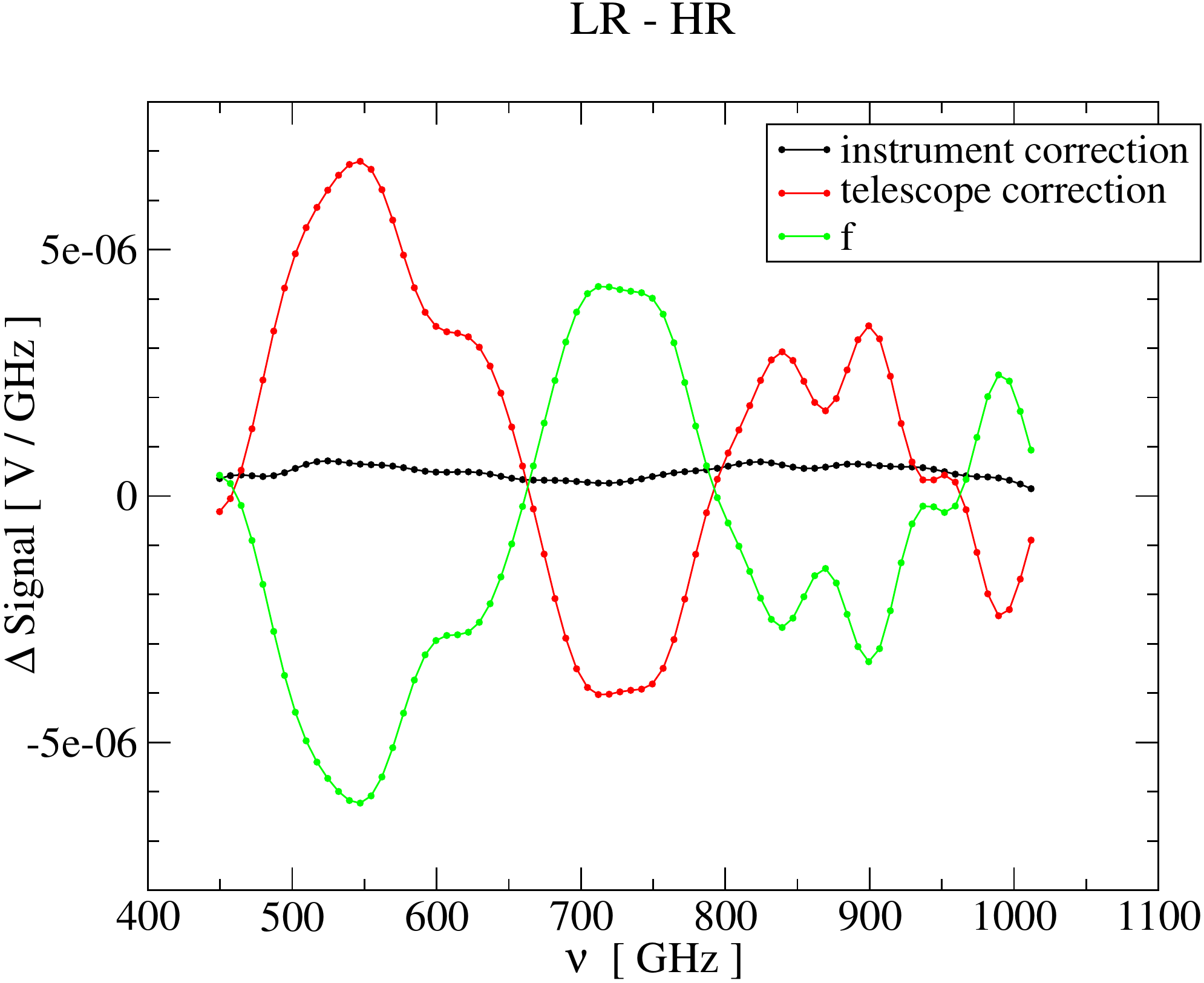}
      \caption{Average differences between the correction curves for LR and HR data.}
              \label{fig:cor-dif}
\end{figure}


This result has two important consequences: i) we regarded $\delta_\mathrm{HR-LR}$ as a constant, but this assumption is correct only to a first approximation; the discrepancies in the response functions at different resolutions imply that the amplitude of the discrepancy must depend on the instrument and telescope temperatures, which vary as a function of time. ii) The existence of $\delta_\mathrm{HR-LR}$ cannot be attributed to systematically wrong estimates of the telescope or instrument temperatures, because no systematic temperature variation could mimic a difference in the RSRFs.

Both conclusions are supported by Fig. \ref{fig:delta_var}, which shows the temporal evolution of $\delta_\mathrm{HR-LR}(\mathrm{ObsID}_\mathrm{HR}, \mathrm{ObsID}_\mathrm{LR})$ for three different frequencies. The 524 and 884 GHz frequencies (black and green dots) correspond to the peaks of the double bump, while 712 GHz (red squares) roughly corresponds to the minimum between the bumps. While the variations at 524 and 884 GHz show correlated trends, the pattern followed by the variations at 712 GHz is approximately antithetical. Therefore the variations of $\delta_\mathrm{HR-LR}(\mathrm{ObsID}_\mathrm{HR}, \mathrm{ObsID}_\mathrm{LR})$ are proportional to $\delta_\mathrm{HR-LR}$, which strongly suggest that the variability is intrinsic to $\delta_\mathrm{HR-LR}$. Also, it would not be possible to simultaneously correct for the variations at 524, 712, and 884 GHz by modifying the telescope or the instrument model, as it would cause coherent signal variations at all frequencies.

\begin{figure}
   \centering
   \includegraphics[width=0.96\columnwidth]{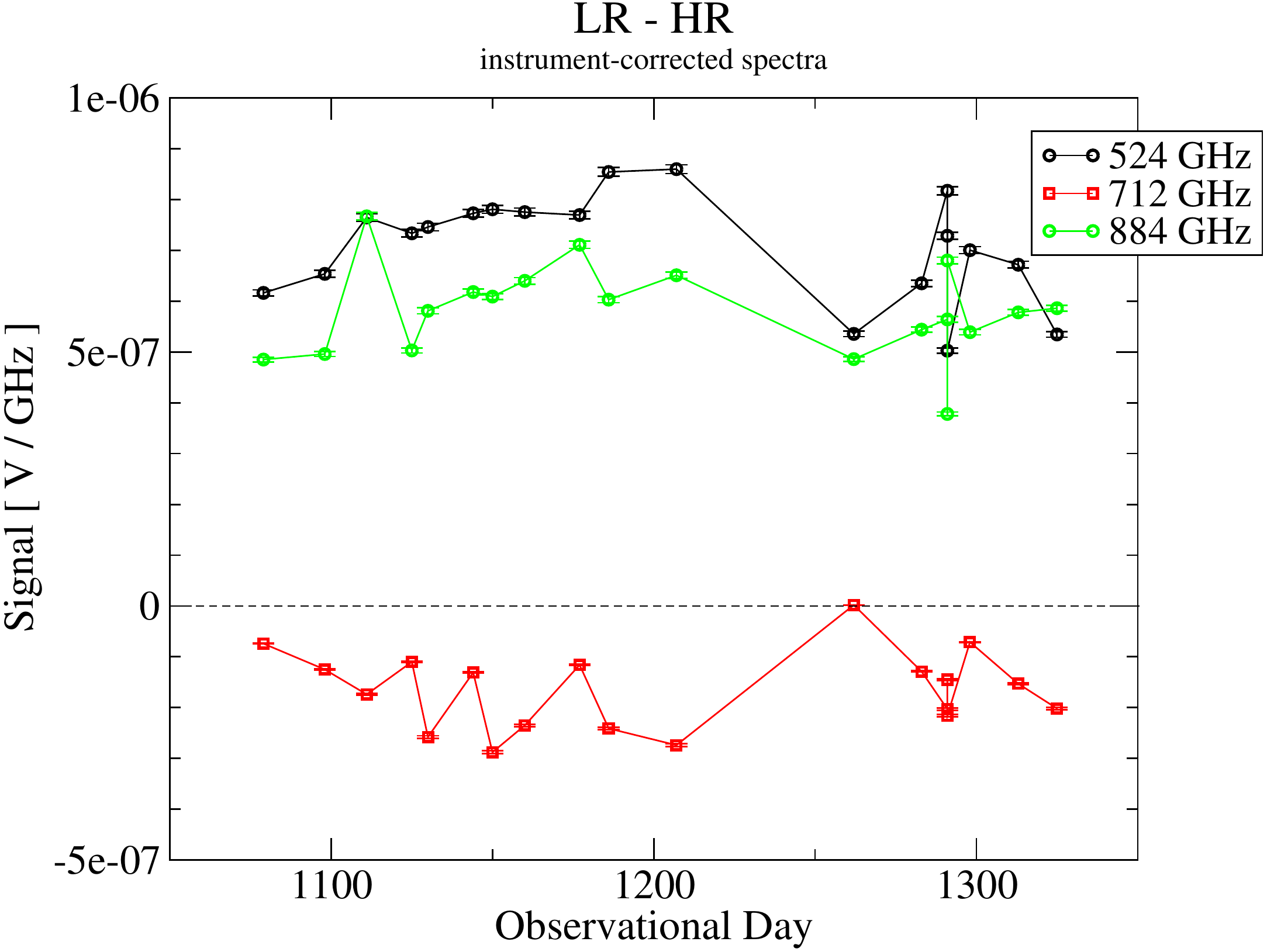}
      \caption{The temporal evolution of $\delta_\mathrm{HR-LR}(\mathrm{ObsID}_\mathrm{HR}, \mathrm{ObsID}_\mathrm{LR})$ at three different frequencies: 524 GHz (black dots), 712 GHz (red squares), and 884 GHz (green dots).}
              \label{fig:delta_var}
\end{figure}

\subsection{Dark sky observations in OD 1291}
\label{Dark}

The different behaviour of the spectral response functions at low and high resolution 
seem to suggest a different sensitivity of the mechanics or the optics to the resolution mode, which is an unlikely scenario. The origin of the problem can be further investigated through the analysis of a set of dark sky observations carried out in OD 1291 and performed in LR, HR, and H+LR resolution modes (see Table \ref{tab:od1291}). The difference between instrument corrected spectra of all the observations from 1342256085 to 1342256089 and the one of obsid 1342256091 are plotted in the lower panel of Fig.  \ref{fig:OD1291}.

\begin{table}
\caption{Dark sky observations carried out in OD 1291 (Col. 1), with their resolution modes (Col. 2) and number of repetitions (Col. 3).\label{tab:od1291}}  
\centering
\begin{tabular}{c c c}
\hline
\noalign{\smallskip}
 Obsid & Resolution mode & Reps\\
 \hline
\noalign{\smallskip}
 1342256085 & LR & 20\\
 1342256086 & LR & 20\\
 1342256087 & HR & 5\\
 1342256088 & LR & 20\\
 1342256089 & H+LR & 25\\
 1342256091 & HR & 70\\
\hline
\end{tabular}
\end{table}

From which several conclusions can be drawn
\begin{itemize}
\item Depending on the temporal sequence in which they are observed, HR and LR spectra are sometimes almost identical (see the green and the blue line). This is consistent with the negligible differences seen between uncalibrated H+LR(H) and H+LR(L) (see Fig. \ref{fig:sdi-fts}).
\item The differences among HR spectra can be similar in shape and comparable in amplitude to $\delta_\mathrm{HR-LR}$ (see, e.g., the green line, showing the difference between obsids 1342256087 and obsid 1342256091);
\item LR spectra show important variations from one observation to another. 
\end{itemize}

All these arguments point towards excluding the resolution mode as the direct cause of the problem. The differences with respect to the 1342256091 spectrum are all characterised by the typical double bump of $\delta_\mathrm{HR-LR}$, but the amplitude of the effect varies considerably. Inspecting the housekeeping products for this set of observations reveals most of the parameters show only mild variation with time --- above all the telescope. However, fast variability can be detected in the instrument temperature. Looking at its behaviour during the data acquisition (upper panel of Fig.  \ref{fig:OD1291}), it is hard to find an unequivocal pattern: i) Most of the spectra seem to gradually tend toward a decreasing difference with respect to 1342256091, which could suggest a time dependence of the amplitude of the effect; however, the transition from obsid 1342256085 (black line) to 1342256086 (red line) seems to go in the opposite direction. ii) Obsid 1342256086 (red line) is characterised by a positive temperature gradient and a larger departure from obsid 1342256091, while the LR spectra in obsids 1342256088 (blue line) and 1342256089 (brown line) by negative temperature gradients and proportionally smaller discrepancies from obsid 1342256091. A correlation between gradients and amplitude of the effect, however, could apply only to LR data, because HR observations seem to contradict it. iii) The most evident distinction between observation 1342256091 (violet line) and all the others is the number of repetitions and the duration of the data acquisition. While the 70 repetitions that comprise 1342256091 allow the instrument temperature to reach an approximately constant value, observations from 1342256086 to 1342256089 show strong temperature gradients; this does not apply to obsid 1342256085 (black line), which is short, but does not show important instrument temperature variations. 

In summary, the discrepancies among the spectra are not connected to the instrument temperature in a straightforward way, while the number of repetitions of the observations seems to have a significant influence on the measured flux densities.

\begin{figure}
   \centering
   \includegraphics[height=1.05\columnwidth, width=0.96\columnwidth]{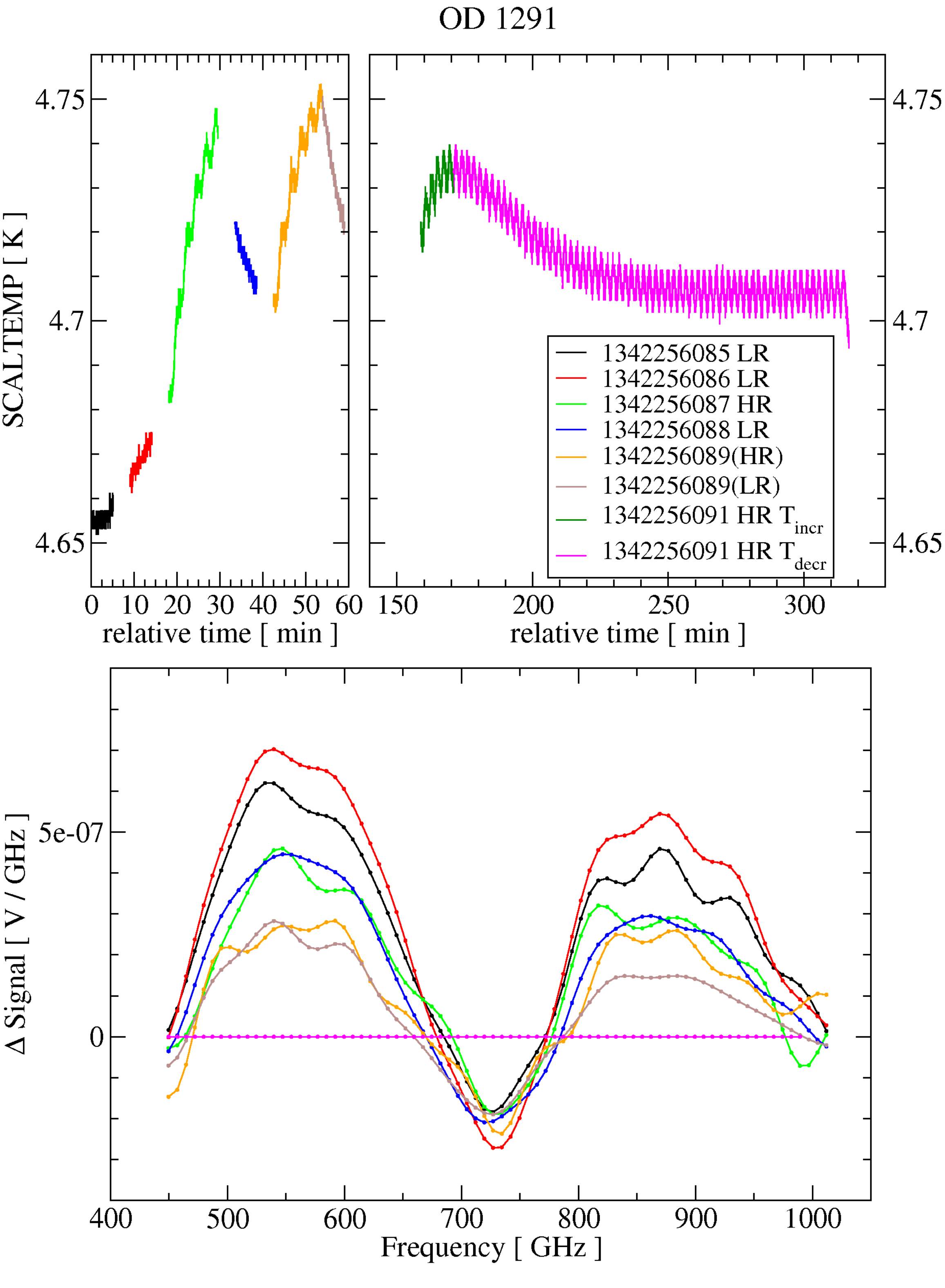}
      \caption{Upper panel: The variation of the instrument temperature, measured by one of the dedicated sensors, during the observations in OD 1291. Lower panel: the spectra of the observations from 1342256085 to 1342256089, after subtracting the spectrum of observation 1342256091. Uncertainties on the plotted values are below 1\%.}
              \label{fig:OD1291}
\end{figure}

\section{A link to the telescope model correction?}
\label{Telescope correction}

The hypothesis of a relationship between number of repetitions of an observation and the amplitude of the discrepancy with respect to an average HR spectrum is particularly interesting when confronted with the result of an independent study about the calibration of the SPIRE FTS data, namely the telescope model correction (see \citealt{Hopwood14}). This study, which takes into account high-resolution spectra, demonstrates the significant improvement of the calibration results after multiplying the telescope model by a time-dependent factor; it is hypothesised that the correction is required because of an extra-emission caused by the build up of dust on the surface of the telescope. It is also shown that the correction factor changes according to the number of repetitions of the observation: observations with $<20$ repetitions require a higher correction factor than those with $>20$ repetitions (see Fig. 5 and 6 in \citealt{Hopwood14}). The authors hypothesise that this difference is caused by a higher than average instrument temperature (most of the short dark sky observations were taken at the end of an FTS observing cycle, when the temperatures are generally higher). However, 
provided that the duration of an observation correlates with the number of repetitions, it might also be that the dependence of the correction factor on the number of repetitions is an indirect consequence of a dependence on the \emph{duration} of the observation. Since LR observations are systematically shorter than the HR ones, it follows that the discrepancy between HR and LR data may be one aspect of a more general problem that has to do with the duration of an observation, rather than its resolution mode.

From the discussion above, one might wonder if the origin of $\delta_\mathrm{HR-LR}$ is the telescope model itself. The multiplication of the original model by a correction factor can have similar effects as a change of the telescope RSRF, which would be compatible with the results reported in Sec. \ref{Three-parameter}, but only up to a point. 
Hypothesising the existence of a correction factor that varies with either the duration or the number of repetitions of an observation, the difference between two spectra could be expressed, as a function of frequency, as $(E_{corr}-{E'_{corr}})M_\mathrm{Tel}(\nu){R_\mathrm{Tel}}(\nu)$, where $E_{corr}$ and $E'_{corr}$ would be the correction functions to apply for the given observations. This difference would be proportional to $M_\mathrm{Tel}(\nu){R_\mathrm{Tel}}(\nu)$, whose shape is known, and is incompatible with the double bump characterising $\delta_\mathrm{HR-LR}$.
Inverting the problem, it could instead be hypothesised that the detected difference between the correction factors is the consequence of observation-duration-dependent RSRFs.

\section{RSRF calculation from scans of the same observation}
\label{Scans}

The differences among the quasi-simultaneous dark sky observations in OD 1291 demonstrate that variation of the RSRFs can occur on very short timescales --- of the order of minutes. A way to investigate such short-term variations is to analyse on a scan-by-scan basis, rather than using the standard pipeline products, which (for each detector) are averaged over all scans per observation. In the following we will focus on the flux densities at a frequency of 524 GHz, around which the discrepancy between LR and HR/H+LR data is most pronounced.

Changes in telescope emission during an observation is minimal and therefore subtraction of a single telescope model per observation is not responsible for any significant deviation of scans from the expected behaviour. In Fig. \ref{fig:planefit_scans}, the telescope corrected signal at 524 GHz from observations 1342256089 (H+LR) and 1342256091 (HR) is plotted versus the instrument model $M_\mathrm{Inst}$. According to Eq. \ref{eq:three-par}, the data-points should fall along a straight line, whose slope should provide the best instrument response function for the set of data, ${R'_\mathrm{Inst}}$(524 GHz). For comparison, the plot also reports the telescope corrected signal at 524 GHz (black dots) and fitted slope (black line) for the LR dark sky (henceforth, LR$_{\mathrm{dark}}$) observations in Table \ref{tab:obsid}.

The HR scans of obsid 1342256089 (orange squares) are the first in order of time and during these ten scans the instrument temperature increases almost monotonically. The data-points for the first three scans are consistent with the average behaviour of low-resolution observations. As the temperature further increases, the decrease in $V_i(\nu)-M_\mathrm{Tel} {R'_\mathrm{Tel}}$ becomes much steeper than expected. This indicates a change of ${R'_\mathrm{Inst}}$(524 GHz) (represented by the slope of the orange line in the figure). Apparently, the response of the detector to the instrument emission has changed, leading to a discrepancy with respect to LR$_{\mathrm{dark}}$ observations. The LR scans of observation 1342256089 were taken directly following the HR part, and are shown as brown squares in Fig. \ref{fig:planefit_scans}. For these, the instrument temperature tends to decrease. The slope of a linear regression for these LR scans (brown line) can be used to infer the related response function, which is similar to the one inferred from the LR$_{\mathrm{dark}}$ observations. This explains why the discrepancy introduced by the high-resolution part of the observation is approximately preserved. Note the alignment between the flux density for the averaged observation (red square) and the fit to these LR data-points.

The scans of the HR obsid 1342256091 are divided in two: the first part comprises the first 10 scans, for which $M_\mathrm{Inst}$ increases (from now on, $1342256091 T_\mathrm{incr}$, green squares), while for the second part of 130 scans, $M_\mathrm{Inst}$ decreases ($1342256091 T_\mathrm{decr}$, magenta squares). The green and magenta arrows show how the data-points move with time across the plot. The starting instrument temperature for observation 1342256091 (start of $1342256091 T_\mathrm{incr}$) is higher compared to that at the start of 1342256089. However the slope (green line) with which $V_i(\nu)-M_\mathrm{Tel} {R'_\mathrm{Tel}}$ decreases with $M_\mathrm{Inst}$ is similar to the one calculated for the high-resolution part of obsid 1342256089, and therefore similarly steeper than for LR$_{\mathrm{dark}}$ observations. The consequence is that the difference of the 524 GHz flux density of 1342256091 increases, with respect to LR$_{\mathrm{dark}}$. It can also be seen that the response function inferred from $1342256091 T_\mathrm{decr}$ (slope of the magenta line) is similar to the one found for LR$_{\mathrm{dark}}$ observations.

\begin{figure}
   \centering
   \includegraphics[width=0.96\columnwidth]{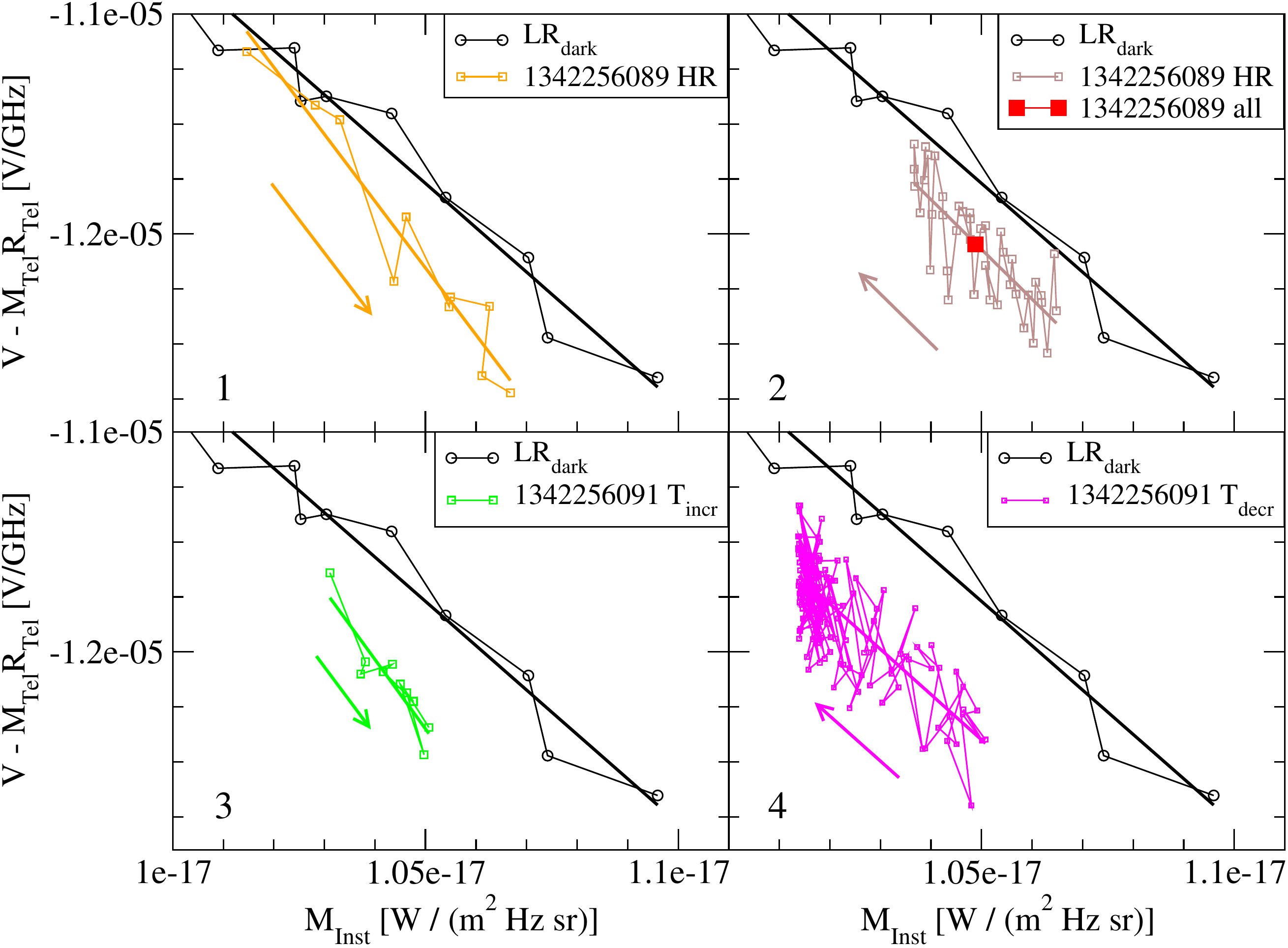}
      \caption{Telescope-corrected signal at 524 GHz plotted versus the instrument model for samples of scans of ObsIDs 1342256089 and 1342256091 (orange, brown, green, and magenta squares); the arrows indicate how the data-points move with time across the plot. In black dots, the telescope-corrected signal of all the LR dark sky observations (LR$_{\mathrm{dark}}$) in Table \ref{tab:obsid}.}
              \label{fig:planefit_scans}
\end{figure}

\begin{table}
\caption{The instrument response functions ${R_\mathrm{Inst}}'$  inferred from different sets of scans of Obsids 1342256089 and 1342256091, compared with the one of all LR observations in Table \ref{tab:obsid}; the sets of scans are selected according to the resolution mode and the monotonic trend of the instrument model variation.\label{tab:scans_slopes}}  
\centering
\begin{tabular}{c c c c}
\hline
\noalign{\smallskip}
 Obsid & Resolution mode & $M_\mathrm{Inst}$ & ${R'_\mathrm{Inst}}$ \\
 \hline
\noalign{\smallskip}
 all LR &  LR  & --- & ($-2.02\pm0.03$)e+12\\
 1342256089 & HR & increasing & ($-3.0\pm0.3$)e+12\\
 1342256089 & LR & decreasing & ($-2.3\pm0.3$)e+12\\
 1342256091 & HR & increasing & ($-3.2\pm0.5$)e+12\\
 1342256091 & HR & decreasing & ($-2.00\pm0.15$)e+12\\
\hline
\end{tabular}
\end{table}

Table \ref{tab:scans_slopes} summarises the main findings. For all scans under consideration (Col. 1), we report the resolution mode (Col. 2), the behaviour of $M_\mathrm{Inst}$ during the scans (Col. 3), and the inferred ${R'_\mathrm{Inst}}$ (Col. 4). These results indicate a link between ${R'_\mathrm{Inst}}$ (which has a direct influence on the measured flux density at 524 GHz) and the variation of $M_\mathrm{Inst}$. The resolution mode does not cause systematic variations of the instrument response function.

The evidence presented so far indicates that slow changes of the instrument temperature cause variation of the voltage density $V_i(\nu)$ (which can be described with the standard formula $M_\mathrm{Inst}(\nu) {R_\mathrm{Inst, LR}}(\nu)$). In contrast, the execution of scans in HR mode initially causes a fast increase of the instrument temperature, which produces a change in the instrument response function and leads to an augmented deviation of the measured flux density when compared to the standard LR observation. 
After a number of HR scans, the instrument temperature stabilises before gradually decreasing. The instrument response function tends to a value slightly steeper than ${R_\mathrm{Inst, LR}}(\nu)$, approximately preserving the flux density difference introduced by the previous scans. Since the absolute value of the response function for these scans 
is marginally higher than ${R_\mathrm{Inst, LR}}(\nu)$, the instrument emission will eventually return to a level that is compatible with $M_\mathrm{Inst}(\nu) {R_\mathrm{Inst, LR}}(\nu)$, although this happens on timescales much longer than the duration of an average observation. As in a hysteresis cycle, the instrument contribution to $V_i(\nu)$ depends on the evolution of the instrument temperature, i.e. the way it changed in previous observations.

\subsection{Extending the scan-by-scan analysis to a large sample of HR observations}
To verify that fast instrument temperature variations are responsible for the problem under discussion, a scan-by-scan analysis was also applied to a sample of 21 HR dark sky observations, from operational day 466 to 1389. They range from 30 to 160 repetitions in length, where one repetition corresponds to two scans.
Since the strongest temperature variations concern the first 20-30 scans of each observation, our analysis is limited to the first 30 and the last 10 scans only. The results are shown in Fig. \ref{fig:planefit_aff} and \ref{fig:planefit_Naff}, where the telescope-corrected signal is plotted versus the instrument model. The green and the orange dotted lines show the typical relationship between signal and instrument model for LR and HR spectra, respectively. Two distinct kinds of behaviour are evident:
\begin{itemize}
\item For the eleven cases shown in Fig. \ref{fig:planefit_aff}, the signal evolves with the changing temperature according to a clearly recognisable pattern. To start with, the signal follows the typical evolution expected for LR spectra (green line). After a number of scans, the signal rapidly moves towards a lower state, which indicates that the contribution of the instrument emission to the signal is higher than expected for the measured temperature. When the instrument temperature starts to decrease, the signal is generally moving along the orange line that characterises HR spectra. By overlapping the patterns followed by the eleven observations under consideration, a model of the resulting \emph{U-shaped} pattern has been obtained (magenta line in the box; its time evolution is clockwise).
\item For the remaining ten observations (see Fig. \ref{fig:planefit_Naff}), the variation of the signal with changing temperature does not follow a clear trend. The signal approximately moves along the typical HR line, without significant deviations.
\end{itemize}

To understand the origin of the differences in behaviour between the two samples of observations, it is useful to place them into the framework of their historical sequence. Eight out of eleven observations in the first group have been performed directly after LR observations, while all the observations in the second group follow HR observations. This result strongly supports the idea that the discrepancies in the calibrated spectra of observations performed in different resolution modes has to do with fast variations of the instrument temperature. When temperature variations are slow (as during or after LR scans), the instrument contribution to the signal follows the typical LR line, while sudden increases of the temperature shift it toward the HR line. If two HR observations are taken in sequence, the system does not have the time to return to the LR state and all the scans will evolve following the typical HR pattern, as illustrated by the example set of 10.

As mentioned in the end of Sec. \ref{Scans}, the instrument emission tends to return to a level compatible with $M_\mathrm{Inst}(\nu) {R_\mathrm{Inst, LR}}(\nu)$ on timescales much longer than an average HR observation. This provides a convincing explanation for the case of the three observations belonging to the first group despite being performed directly after HR spectra. Two of them precede intermediate mode HR observations, whose duration is generally more than three times longer than for sparse mode. The third observation follows instead a sequence of sparse HR observations lasting about 11 hours.

\begin{figure}
   \centering
   \includegraphics[width=0.96\columnwidth]{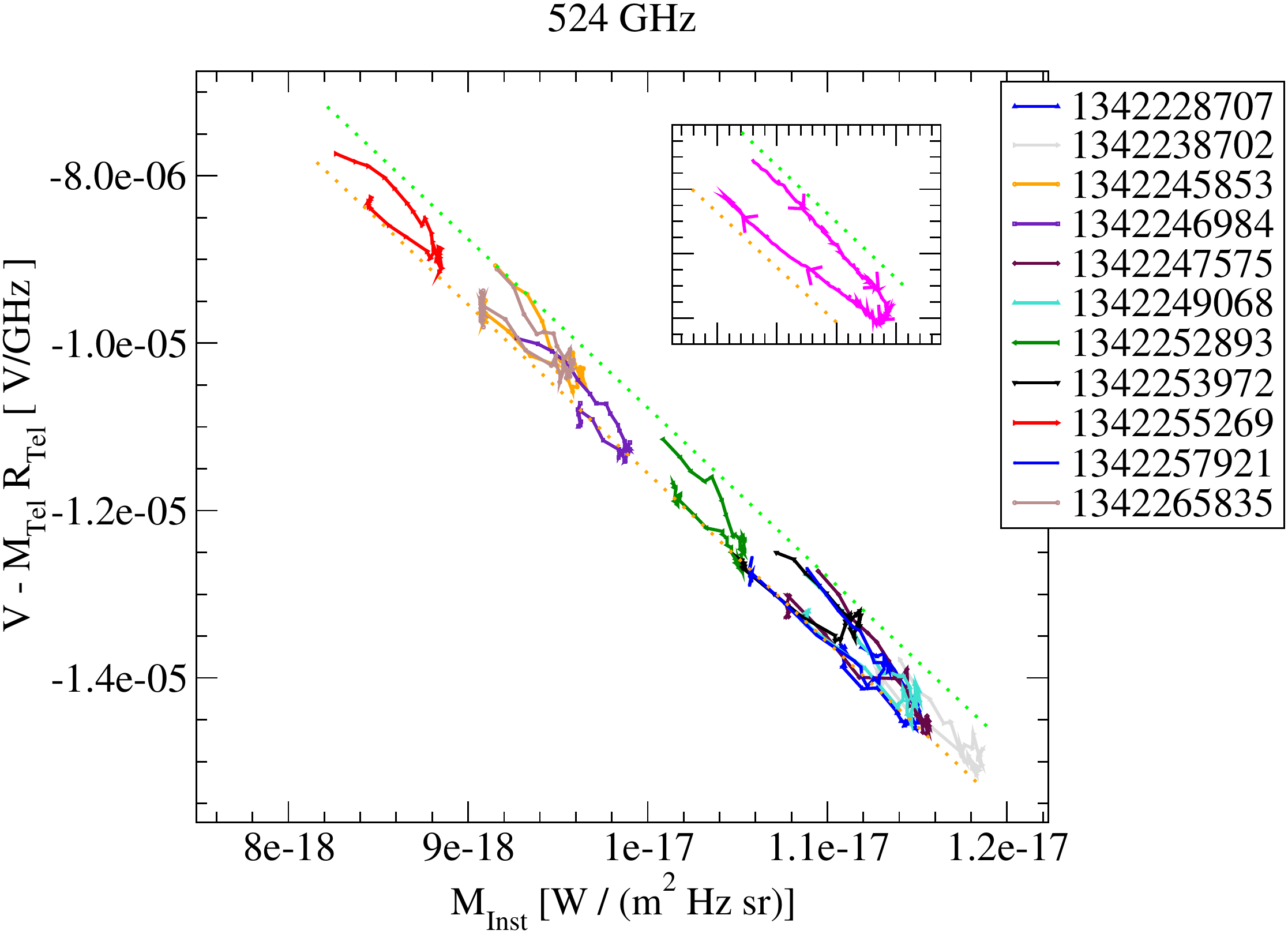}
      \caption{Telescope-corrected signal at 524 GHz plotted versus the instrument model for samples of scans of HR observations. The signal follows a kind of U-shaped pattern (a model of this pattern is shown as a magenta line in the top-right box), moving clockwise from a typical LR behavior (green dotted line) to a typical HR one (orange dotted line).}
              \label{fig:planefit_aff}
\end{figure}
\begin{figure}
   \centering
   \includegraphics[width=0.96\columnwidth]{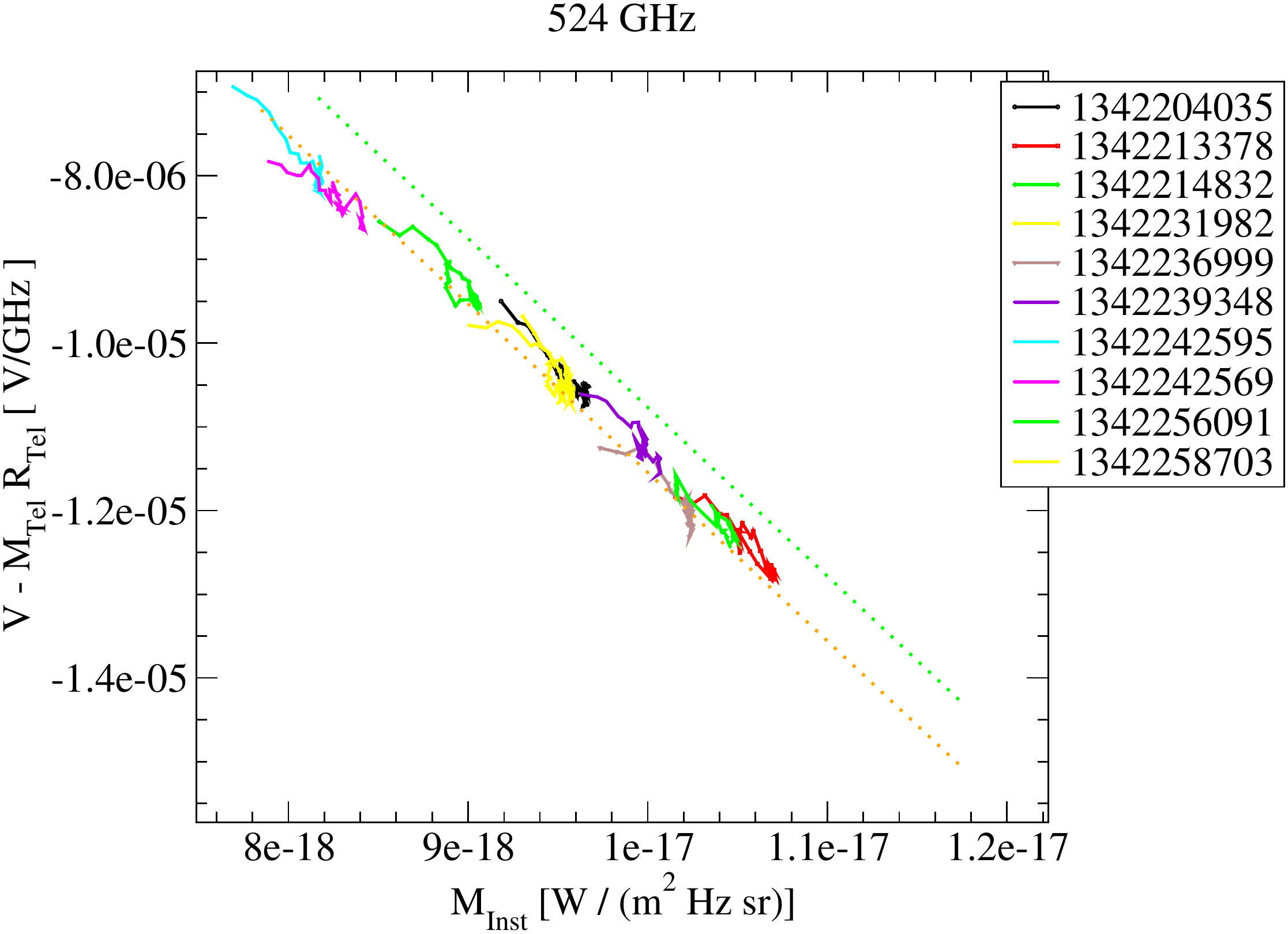}
      \caption{The evolution of the telescope-corrected signal at 524 GHz for a different group of HR observations. All the scans seem approximately to move along the typical HR line.}
              \label{fig:planefit_Naff}
\end{figure}

The scenario above provides a coherent description of the variations observed in a large sample of dark sky spectra, although it does not explain the origin of the problem. Comparing the temperatures reported by the three sensors placed in different positions within the instrument (\textsc{scalTemp}, \textsc{scalTemp2}, and \textsc{scalTemp4}) we see a significant delay between the variations triggered by the first scans of each high-resolution observation. It could be an indication that the problem has to do with the thermal balance of the system; the question to address, then, is whether thermal balance can influence the response functions.

\section{Variations of the telescope response function}


The analysis illustrated in the previous sections revealed several interesting aspects concerning the systematic discrepancy between LR and HR/H+LR data, although 
theses clues do not fully converge towards an unambiguous definition of the problem's origin.

The separate calculation of the RSRFs for HR and LR observations (see Section \ref{Three-parameter}) shows that changes of resolution mode trigger stronger variations in the telescope contribution to $V_i(\nu)$ than in the instrument contribution (see Fig. \ref{fig:cor-dif}), with the frequency dependent variation of the former being consistent with the double-bump of $\delta_\mathrm{HR-LR}$. The dependence of the telescope correction factor on the number of repetitions could suggest a link with the observation-duration dependent variation of the telescope RSRF hypothesised in Section \ref{Telescope correction}. All these points would argue in favour of a strong involvement of the telescope emission in the flux density discrepancies. On the other hand, the telescope temperature seems to be nearly constant on timescales of minutes to a few hours, excluding its strong involvement in the variations of the spectra acquired in OD 1291.

The most likely scenario seems to be a change of the response functions triggered by the instrument temperature: since HR observations are characterised by temperatures that are generally higher than those of LR observations, the systematic nature of the discrepancy would be explained. Assuming that fast changes of the instrument temperature can affect both the telescope and the instrument response functions, the analogies found with some characteristics of the telescope correction would also be justified. 

\section{Data correction}
\label{correction}

Although the origin of the presented problem has been identified, there is no obvious way to analytically correct the data. The amplitude of the spurious signal distorting the shape of the continuum in LR spectra depends on both the sequence and characteristics of previous observations, which cannot be described by simple combinations of housekeeping parameters. The situation is complicated further by the fact that the response functions are empirically calculated in a way that tends to minimise the residual noise, which means that part of the spurious signal may be removed by the standard calibration.

The correction of the spectra can be achieved empirically, by exploiting the peculiar shape of the systematic noise introduced by the variation of the response functions (see Fig. \ref{fig:corr}). The characteristic double-bump observed in the continuum of the affected observations varies in amplitude, however its shape is approximately constant, and can be precisely modelled.

\begin{figure}
   \includegraphics[width=0.96\columnwidth]{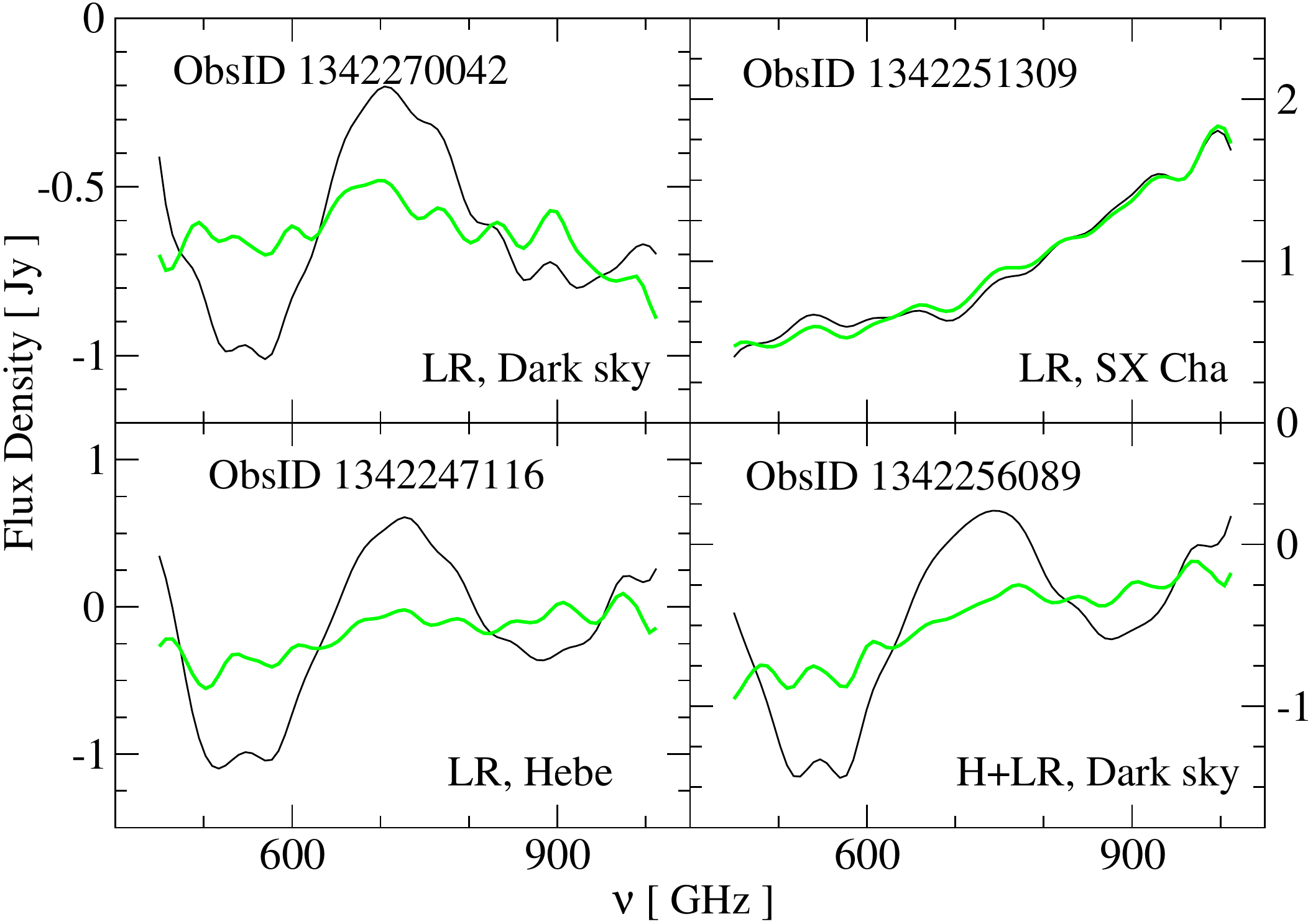}
      \caption{Low resolution spectra before (black line) and after (green line) the empirical correction of the data. The correction efficiently corrects both H+LR(L) (bottom right panel) and LR (all other panels) observations. When no double-bump is identifiable in the data (top right panel) the changes introduced by the correction are negligible.}
              \label{fig:corr}
\end{figure}

An algorithm for the a-posteriori correction of low-resolution spectra, both from LR and H+LR(L) observations, has been developed, based on the cross-correlation of an archetype of $\delta_\mathrm{HR-LR}$ with the spectrum to correct. The $\delta_\mathrm{HR-LR}$ archetype (from now on, $\overline{\delta}_\mathrm{HR-LR}(\nu)$) has been calculated by selecting a set of quasi-simultaneous HR and LR spectra, whose differences are clearly affected by the LR calibration problem. These differences have been scaled in order to have the same standard deviation, and then averaged to provide a robust model. 

Given a calibrated spectrum S$(\nu)$, with frequency-averaged value $\bar{\mathrm{S}}$, the corrected spectrum S$'(\nu)$ can be expressed as

\begin{equation}
\mathrm{S}'(\nu)=\mathrm{S}(\nu)-\frac{\sum_{i} \big((\mathrm{S}(\nu_i)-\bar{\mathrm{S}}) \cdot \overline{\delta}_\mathrm{HR-LR}(\nu_i)\big)}{\sigma_{\overline{\delta}}^2} \cdot \overline{\delta}_\mathrm{HR-LR}(\nu)
\end{equation}
where $\sigma_{\overline{\delta}}$ is the standard deviation of $\overline{\delta}_\mathrm{HR-LR}(\nu)$. This is equivalent to calculating the amplitude of the spurious noise for S$(\nu)$ and removing it, assuming its shape to be $\overline{\delta}_\mathrm{HR-LR}(\nu)$.

When a calibrated spectrum has a strong continuum, the amplitude of the spurious noise may be overestimated. This overestimate can be avoided by fitting and subtracting a second-order polynomial to remove the continuum before calculating the noise amplitude. Such a subtraction is only necessary for sources where the continuum level at 1000 GHz is found to be greater than 0.5 Jy.

The LR correction developed is scaled depending on the amplitude of the double-bump present. This means that if the double-bumps are negligible, then the corrections applied are negligible too, and the data are essentially unchanged. It is only applied to SLW detectors, which are the only ones affected by the problem, and efficiently corrects both LR and H+LR(L) observations.

The correction was developed for point-source calibrated spectra, using sparse-mode observations. The standard FTS pipeline also provides extended-source calibrated data, which are either spectra for sparse-mode observations, or spectra and spectral cubes for intermediate and fully sampled mapping observations (see \citealt{Fulton16} for full details of the FTS data reduction pipeline). The LR correction is not applied directly to the extended-source calibrated data, but is propagated by reversing the point-source conversion once the point-source calibrated spectra have been corrected. For LR mapping observations, the conversion to point-source calibrated is an additional step for the SLW array, purely so the double-bump correction can be applied. Fig. \ref{fig:corrExt} shows an example of the extended-source calibrated spectra from the centre detectors for observation 1342243638, before and after correction for the double bump. It should be noted that there is no point-source conversion factor for the vignetted detectors. These detectors are removed before point-source calibration and therefore remain uncorrected in the final extended-source calibrated products. For sparse observations, the uncorrected spectra from these detectors should only be used with caution. For mapping observations, the spectra for all SLW detectors are present in the pre-processed cube, which collates the spectra ready for gridding into a spectral cube. However the spectra from the vignetted detectors are not included in those used to create the standard pipeline SLW cube products.

\begin{figure}
   \includegraphics[width=0.96\columnwidth]{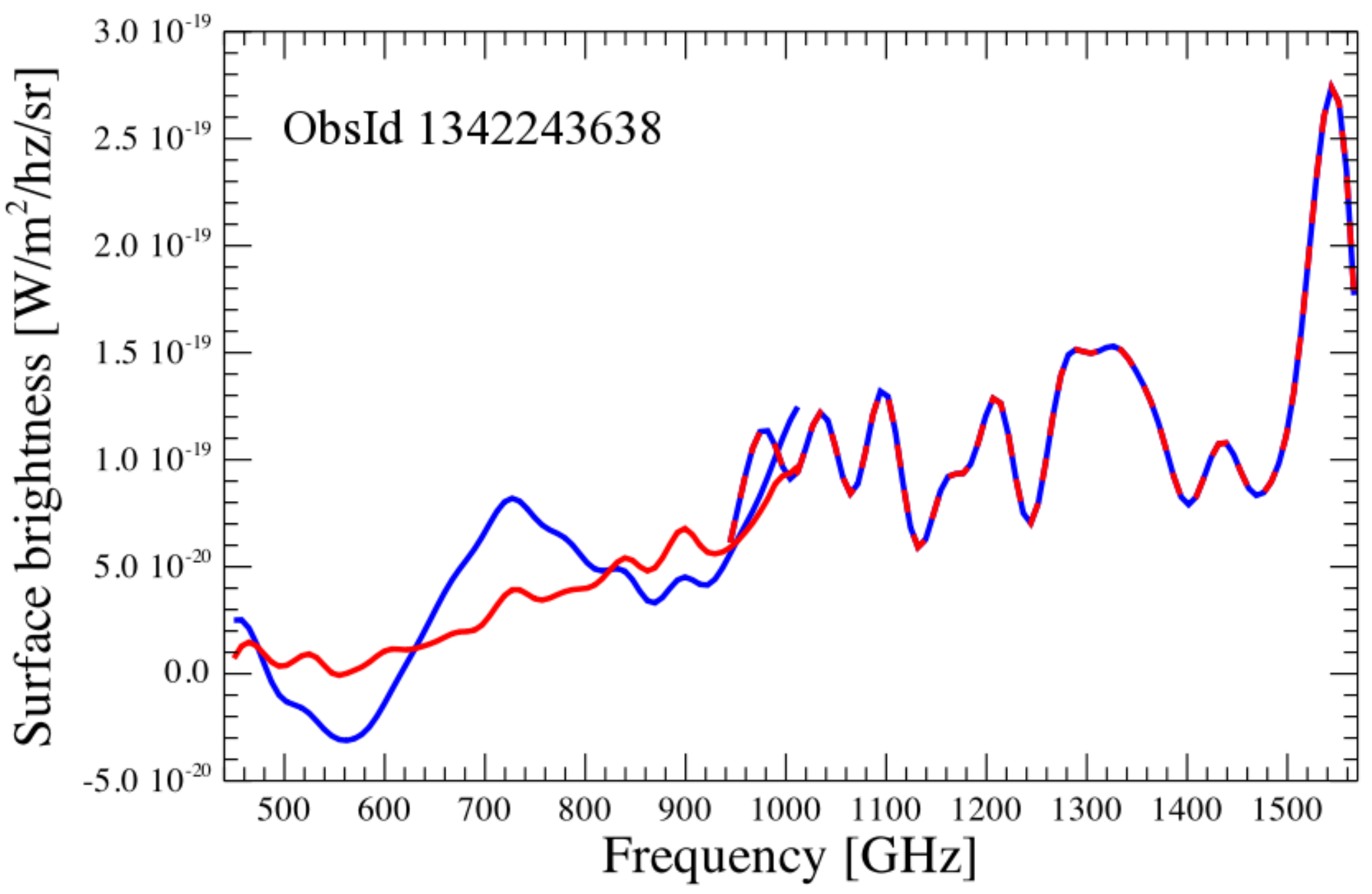}
      \caption{Extended-source calibrated spectra for observation 1342243638. The spectrum before the correction for the double bump is shown in blue, while the one after the correction is shown in red.}
              \label{fig:corrExt}
\end{figure}

\section{LR calibration uncertainty}

Due to the pronounced systematic noise the derived correction removes from the SLW spectra of LR observations, here we report the LR calibration uncertainties, before and after the correction has been applied.
LR sensitivity is assessed following the method in \cite{Hopwood15}. Briefly, the spectral noise within 50 GHz frequency bins is measured for 34 LR observations of the SPIRE dark sky field. The noise taken for each dark sky observation is scaled to 1 $\sigma$ in 1 hour (i.e. the sensitivity) and then the median is found over all the observations for each frequency bin.
There is negligible change in the sensitivity after the LR correction has been applied to the centre SLW detector (see Fig. \ref{fig:uncert1}), which is expected. However the correction of large-scale distortion in the continuum does significantly affect the uncertainty associated with continuum measurements. Fig. \ref{fig:uncert2} shows this "continuum offset" (1 $\sigma$ additive uncertainty), before and after the LR correction of dark sky. Again, the method detailed in \cite{Hopwood15} is followed. The same set of LR dark sky observations are used as for the sensitivity, but in their un-averaged form, i.e. as individual scans. The spectra are smoothed to remove small scale noise and then the standard deviation over all the scans of the 34 dark sky is taken to provide the continuum offset. Before the correction is applied, the effect of the double bump is clearly evident, and when compared to the HR continuum offset (Fig. 32 in \citealt{Hopwood15}), the shape of the curve is notably different. This indicates that although HR data can suffer from the same "bumpiness" as LR data (but generally as an inverted double bump), overall this is not a significant issue for HR, as the HR continuum offset is a relatively smooth curve without any correction for this systematic effect. After the bumpiness has been removed from the LR dark sky, the resulting continuum offset shows an improved and more similar form as to that seen for HR. Figure \ref{fig:LRoffs1} (in the Appendix) shows the continuum offset for all unvignetted SLW detectors, indicating there is a range in how the curves differ when comparing the continuum offset found before and after the LR correction is applied. For some detectors there is almost no change, while for others there is a pronounced improvement. These figures illustrate that the double bump affects different detectors to differing degrees and the correction is working harder where needed.
Similar results are found for the extended-calibrated sensitivity and continuum offset, although the improvement for SLW is somewhat easier to see for the point-source calibrated data, as the difference in beam size across the frequency array has been accounted for.
The average results for both FTS calibrations schemes, and for all off-axis detectors, can be found in Tab. \ref{tab:offsetSensitivity}. The sensitivity is relatively flat, so the mean sum is used for the average. Similarly to the case of SLWC3 (see Fig. \ref{fig:uncert1}), the correction of the double-bump in off-axis detectors leads to marginal differences in the sensitivity. The continuum offset is also relatively flat, but at the centre of the SLW and SSW bands, and only once the LR correction has been applied. However the high systematic noise at the edge of the bands causes the offset to rise sharply even for the corrected data. Therefore, for the average values, the median is taken over 500-900 GHz for SLW and over 1000-1500 GHz for SSW. Note that as there is no LR correction for the vignetted detectors (as discussed in Section 7), only the uncorrected results for these detectors are presented. However, all detectors are included in Tab. \ref{tab:offsetSensitivity}. The extended-source calibrated data used does include the correction applied for the feed-horn coupling efficiency, as described in Valtchanov et al. in prep., and is a comparison of HIPE version 14.0 and HIPE version 14.1. Whereas, for the point-source calibrated data, the LR correction was already in place for HIPE version 14.0 and so this is compared to HIPE version 13.0 reduced data.

\begin{figure}
   \includegraphics[width=0.96\columnwidth]{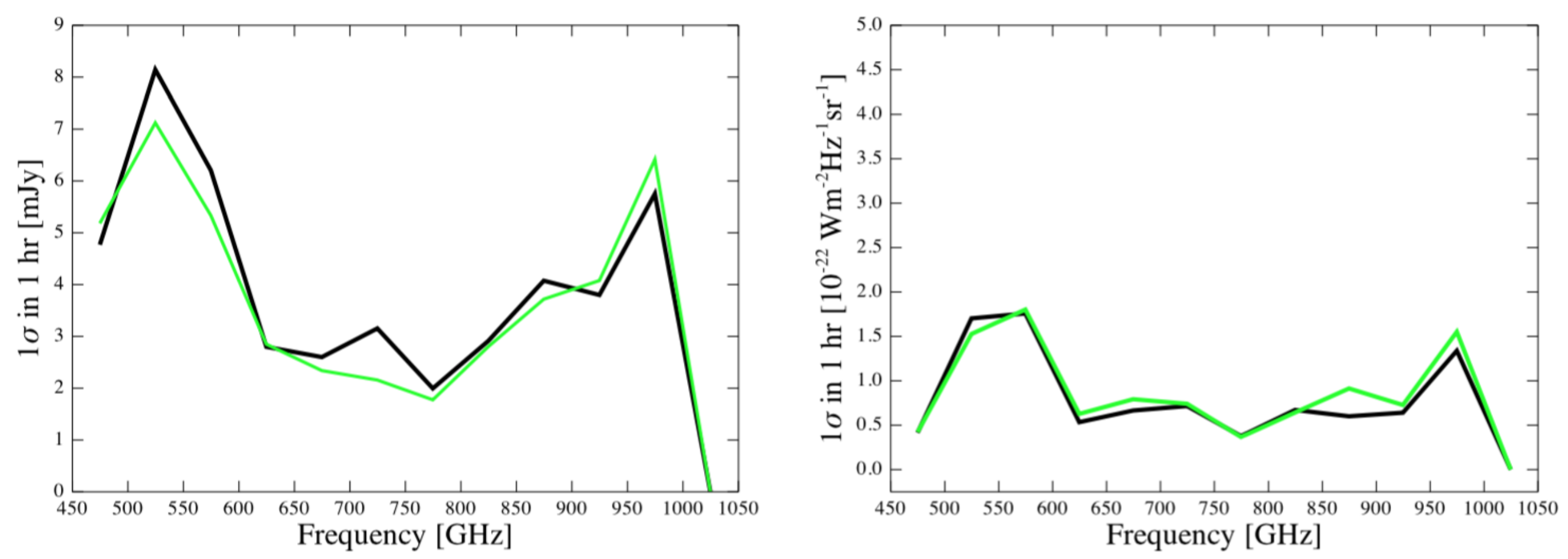}
      \caption{LR sensitivity for point-source calibrated SLWC3 spectra on the left and extended- source calibrated SLWC3 spectra on the right. Black curves show the sensitivity before the application of the LR correction, which is compared to after LR correction in green. The other centre detector (SSWD4) is not shown as the LR correction does not apply.}
              \label{fig:uncert1}
\end{figure}

\begin{figure}
   \includegraphics[width=0.96\columnwidth]{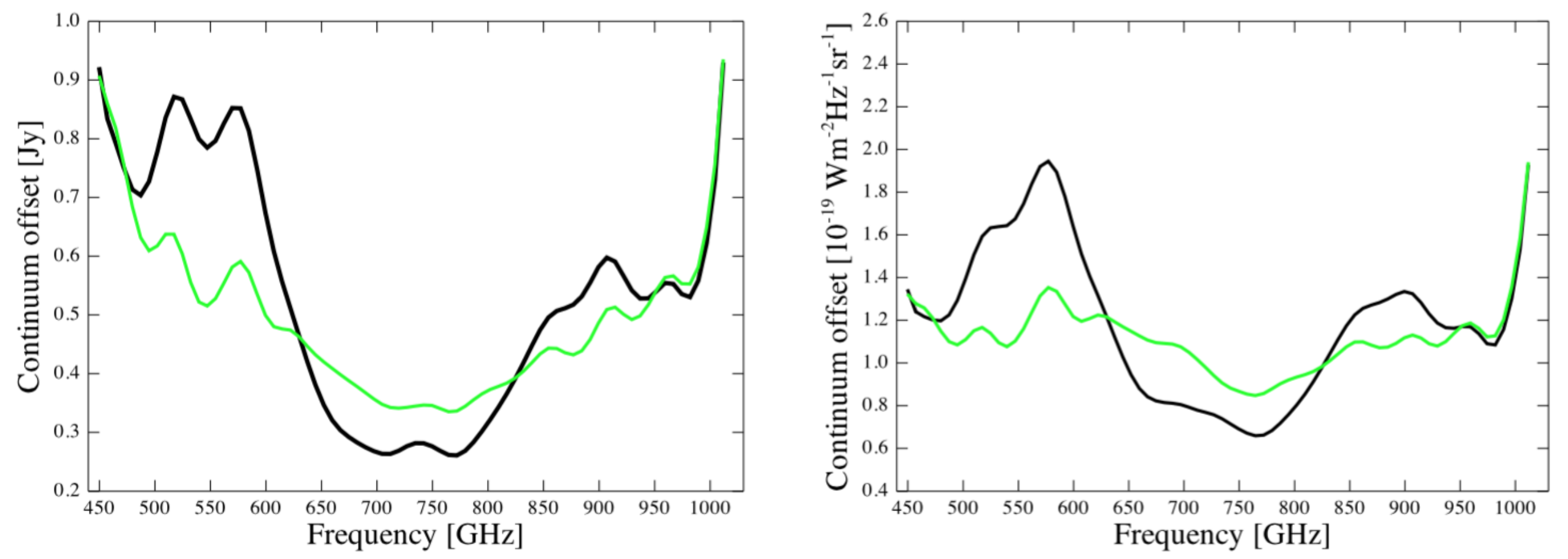}
      \caption{LR continuum offset for point-source calibrated SLWC3 spectra on the left and extended-source calibrated SLWC3 spectra on the right. Black curves show the continuum offset before the application of the LR correction, which is compared to after LR correction in green. The other centre detector (SSWD4) is not shown as the LR correction does not apply.}
              \label{fig:uncert2}
\end{figure}

\begin{table*}
\scriptsize
\caption{Average continuum offset (Offset) and 1\,$\sigma$ in 1 hour sensitivity ($\sigma$). A subscript of PS indicates results for point-source calibrated data. A subscript of EXT indicated results for extended-source calibrated data in units of 10$^{-22}$ Wm$^{-2}$Hz$^{-1}$sr$^{-1}$ for sensitivity and 10$^{-19}$ Wm$^{-2}$Hz$^{-1}$sr$^{-1}$ for the continuum offset. The values for the centre detectors are in the top two rows and shown in bold. No PS value is given for vignetted detectors, as there is no point-source conversion factor for these.}
\medskip
\begin{center}
\begin{tabular}{ccccc}
\hline\hline
Detector & Offset$_{\rm EXT}$ & Offset$_{\rm PS}$[Jy] & $\sigma_{\rm EXT}$ & $\sigma_{\rm PS}$ [mJy] \\ \hline
{\bf SLWC3} & {\bf 1.1048} & {\bf 0.4678} & {\bf 0.7342} & {\bf 3.2842} \\
{\bf SSWD4} & {\bf 3.2032} & {\bf 0.3816} & {\bf 2.9040} & {\bf 4.0326} \\
SSWE4 & 3.1247 & 0.3747 & 3.0639 & 3.7165 \\
SSWE3 & 3.0914 & 0.3591 & 2.4319 & 3.3576 \\
SSWD3 & 3.1208 & 0.3646 & 2.7709 & 3.1690 \\
SSWC3 & 3.4428 & 0.3966 & 2.7952 & 2.9142 \\
SSWC4 & 3.5807 & 0.4249 & 2.1750 & 3.2900 \\
SLWD2 & 1.0569 & 0.4462 & 0.9077 & 3.2511 \\
SLWD3 & 1.4460 & 0.6373 & 0.8577 & 2.9710 \\
SLWC4 & 1.2151 & 0.5478 & 0.8629 & 3.8282 \\
SLWB2 & 1.7839 & 0.8264 & 0.6832 & 3.6886 \\
SLWB3 & 0.9766 & 0.4722 & 0.7133 & 3.4030 \\
SLWC2 & 1.5947 & 0.7129 & 0.7292 & 3.3853 \\
SSWB2 & 3.2576 & 0.3913 & 2.4471 & 2.7915 \\
SSWB4 & 5.9362 & 0.6831 & 4.2169 & 5.7311 \\
SSWD2 & 3.3839 & 0.3962 & 3.5350 & 3.4259 \\
SSWD6 & 3.4571 & 0.4102 & 2.3246 & 3.0039 \\
SSWF2 & 3.9843 & 0.4698 & 3.2448 & 3.3806 \\
SSWE2 & 4.2563 & 0.4945 & 3.3068 & 3.4744 \\
SSWE5 & 3.5452 & 0.4284 & 2.5264 & 3.2531 \\
SSWF3 & 3.6701 & 0.4501 & 2.6771 & 3.0600 \\
SSWC5 & 3.7796 & 0.4215 & 2.8246 & 2.9173 \\
SSWC2 & 3.1391 & 0.3754 & 3.0911 & 3.3979 \\
SSWB3 & 3.7225 & 0.4279 & 3.3304 & 3.7956 \\
SLWA1 & 7.0461 & --- & 1.3062 & --- \\
SLWA2 & 2.8993 & --- & 0.6096 & --- \\
SLWA3 & 4.0916 & --- & 1.1196 & --- \\
SLWB1 & 5.4564 & --- & 1.4960 & --- \\
SLWB4 & 1.7080 & --- & 0.8565 & --- \\
SLWC1 & 7.6510 & --- & 1.4398 & --- \\
SLWC5 & 3.9158 & --- & 1.4591 & --- \\
SLWD1 & 2.3934 & --- & 0.6105 & --- \\
SLWD4 & 1.5106 & --- & 1.3803 & --- \\
SLWE1 & 2.1612 & --- & 0.9928 & --- \\
SLWE2 & 1.0859 & --- & 0.7256 & --- \\
SLWE3 & 1.8117 & --- & 0.8180 & --- \\
SSWA1 & 3.7088 & --- & 2.6783 & --- \\
SSWA2 & 3.2954 & --- & 3.0671 & --- \\
SSWA3 & 3.6980 & --- & 3.0089 & --- \\
SSWA4 & 3.4473 & --- & 2.3017 & --- \\
SSWB1 & 3.1420 & --- & 2.6001 & --- \\
SSWB5 & 3.3644 & --- & 2.2413 & --- \\
SSWC1 & 3.4709 & --- & 2.8922 & --- \\
SSWC6 & 3.7530 & --- & 2.5802 & --- \\
SSWD1 & 3.7946 & --- & 2.2243 & --- \\
SSWD7 & 4.1899 & --- & 5.6228 & --- \\
SSWE1 & 3.3072 & --- & 2.6670 & --- \\
SSWE6 & 3.8144 & --- & 2.3645 & --- \\
SSWF1 & 3.2737 & --- & 2.7927 & --- \\
SSWF5 & 3.9390 & --- & 2.8959 & --- \\
SSWG1 & 3.7410 & --- & 3.1032 & --- \\
SSWG2 & 3.4729 & --- & 3.1657 & --- \\
SSWG3 & 4.3190 & --- & 3.3313 & --- \\
SSWG4 & 5.4466 & --- & 4.1596 & --- \\ \hline
\hline

\end{tabular}
\end{center}
\label{tab:offsetSensitivity}
\end{table*}

\section{Summary}

In this paper we present a thorough analysis of the systematic discrepancy between spectra obtained at different resolution modes. The discrepancy, which was first detected in the fully calibrated H+LR(H) and H+LR(L) spectra, originates in a systematic difference between uncalibrated LR and HR/H+LR spectra, which the standard calibration pipeline partially corrects by assuming different response functions for LR and HR data.

We tested the hypothesis that the discrepancy, $\delta_\mathrm{HR-LR}$, is constant in time, and therefore whether it can be corrected by adding a constant parameter to the standard two-parameter calibration pipeline. The test showed that, in first approximation, the amplitude of the discrepancy can be regarded as constant; 
however, a minor part of the discrepancy is certainly affected by the (time-dependent) telescope and instrument temperatures. Given the shape of $\delta_\mathrm{HR-LR}$ (which shows an excess of signal around 550 and 900 GHz, and a deficiency of signal around 700 GHz), a correction of the problem can not be achieved by modifying the telescope and/or the instrument model; a modification of the response functions would be needed.

The analysis of the variations in dark sky observations spectra performed in OD 1291, even on a scan-by-scan level, seems to indicate that the amplitude of the discrepancy is {\it not directly} related to the instrument temperature or its increasing/decreasing trend. Rather, the amplitude appears to be affected by the history of the temperature variations before the start of the observation. The extension of the scan-by-scan analysis to a sample of 21 HR observations strongly supports this conclusion.

We hypothesise that the existence of $\delta_\mathrm{HR-LR}$ is related to fast temperature changes in the instrument, which cause temporary variations of both the telescope and the instrument response functions. The sensitivity of the telescope response function to the instrument temperature may also explain the important analogies observed between some characteristics of $\delta_\mathrm{HR-LR}$ and the telescope emission, such as the dependence on the number of repetitions (and consequently the duration) of an observation.

Given that the housekeeping parameters do not trace the SLW LR calibration problem in all its aspects, an analytical correction is not possible. Instead, an empirical a-posteriori correction has been developed, based on the cross-correlation of a calibrated spectrum with the characteristic double-bump the problem causes. The strength of the correction applied is dependent only on the  amplitude of the bumps present. With this empirical method it has been possible to significantly reduce the spectral artefacts that appear in the long wavelength spectrometer channel, for both LR spectra and the LR spectra of H+LR observations.

The analysis of the LR calibration uncertainties shows that applying the correction results in an efficient removal of the double-bump. This is true for even the most pronounced cases, while for the rare cases where the double bump feature is negligible the spectra remains essentially unchanged.

\section*{Acknowledgements}
\emph{Herschel} is an ESA space observatory with science instruments provided by European-led Principal Investigator consortia and with important participation from NASA. SPIRE has been developed by a consortium of institutes led by Cardiff Univ. (UK) and including: Univ. Lethbridge (Canada); NAOC (China); CEA, LAM (France); IFSI, Univ. Padua (Italy); IAC (Spain); Stockholm Observatory (Sweden); Imperial College London, RAL, UCL-MSSL, UKATC, Univ. Sussex (UK); and Caltech, JPL, NHSC, Univ. Colorado (USA). This development has been supported by national fund- ing agencies: CSA (Canada); NAOC (China); CEA, CNES, CNRS (France); ASI (Italy); MCINN (Spain); SNSB (Sweden); STFC, UKSA (UK); and NASA (USA). N.M.'s research activity is supported by the VIALACTEA Project, a Collaborative Project under Framework Programme 7 of the European Union funded under Contract \#607380, that is hereby acknowledged.

\appendix
\section{LR continuum offset}
Here below, the plots of the continuum offset for all unvignetted SLW detectors are shown. Note that the correction of the spurious double-bump in LR spectra has only been applied to these detectors.


\begin{figure*}
\begin{center}
\makebox{
\includegraphics[width=0.25\hsize]{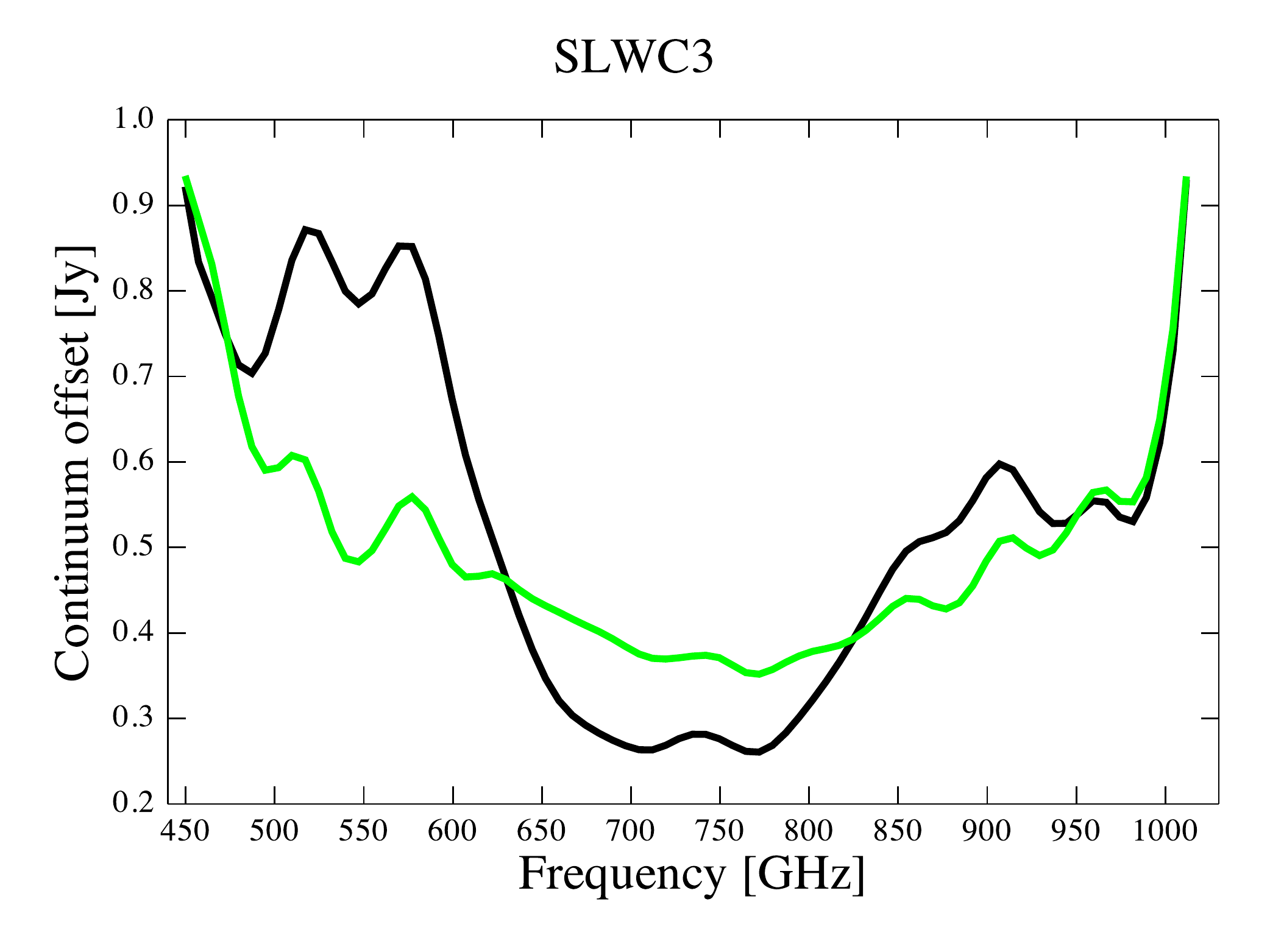}
\includegraphics[width=0.25\hsize]{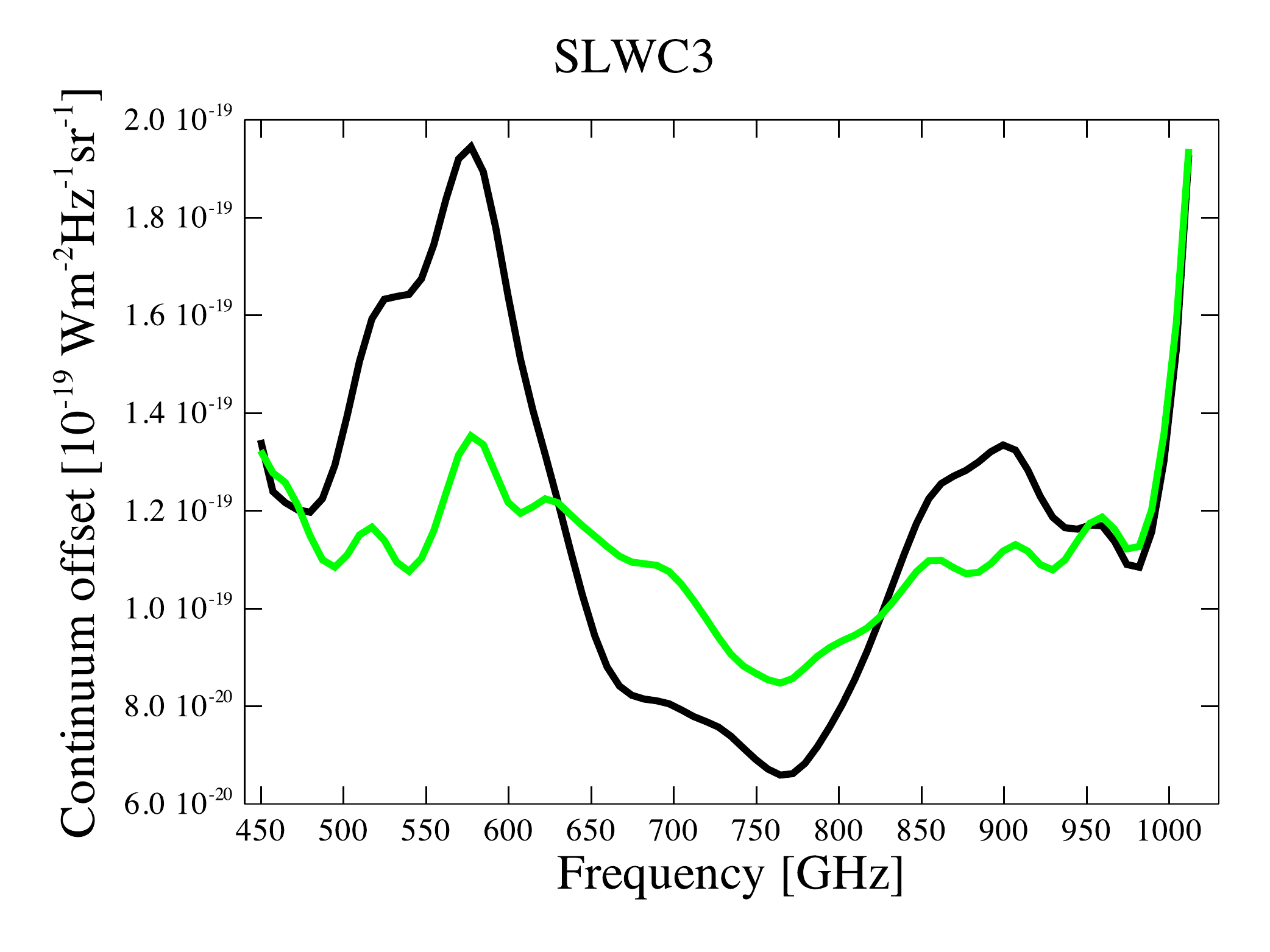}
\includegraphics[width=0.25\hsize]{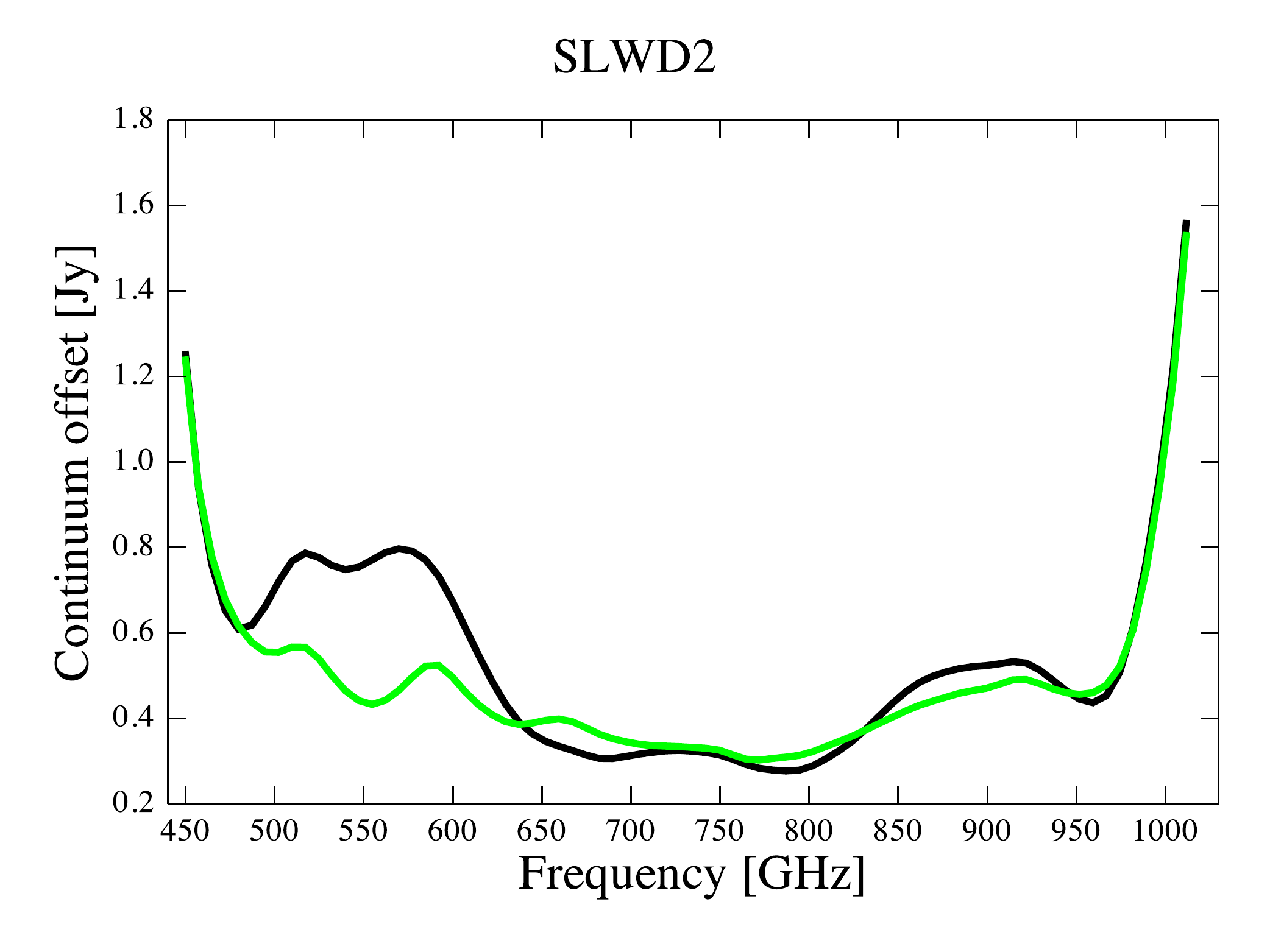}
\includegraphics[width=0.25\hsize]{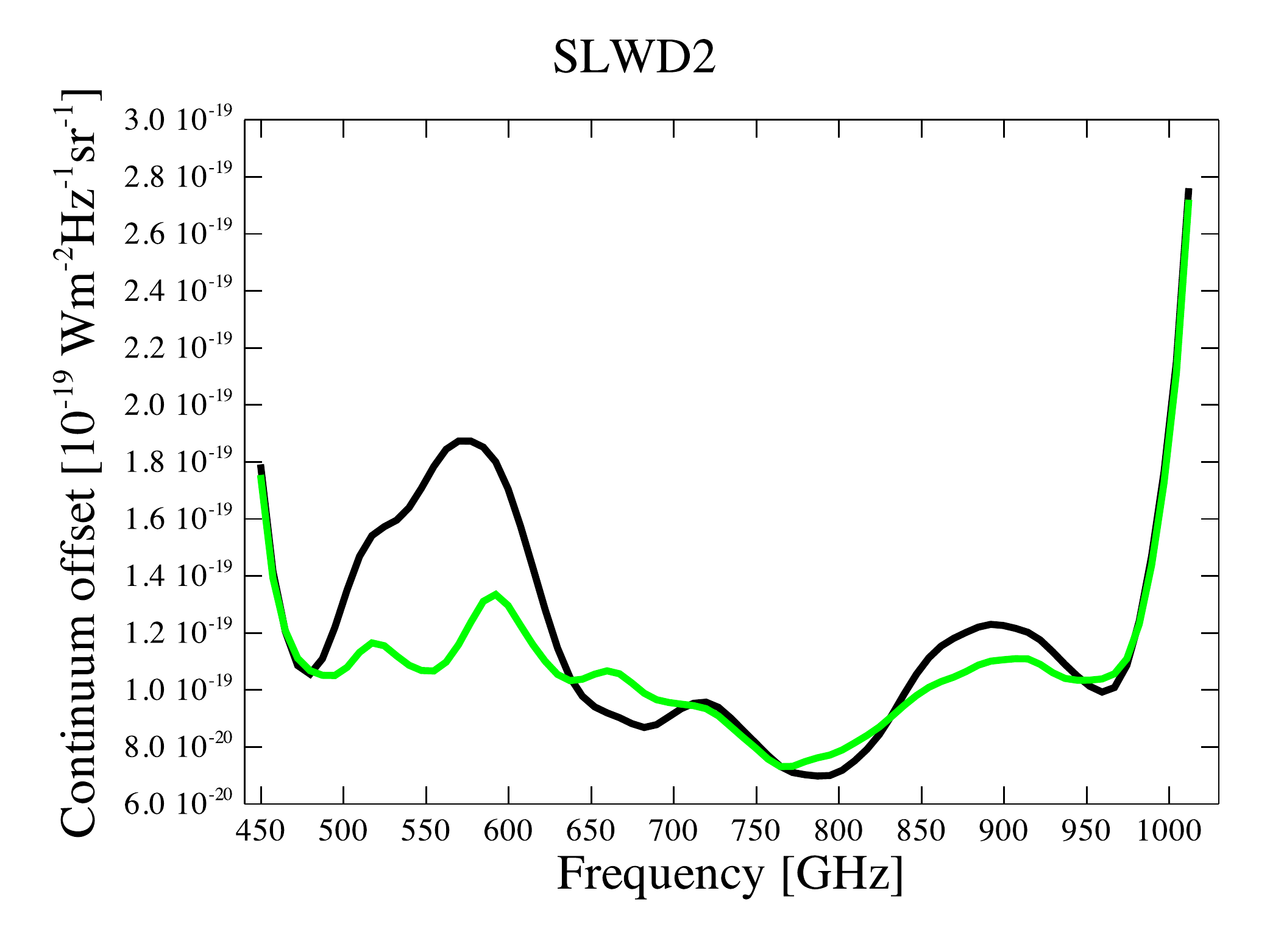}
}
\makebox{
\includegraphics[width=0.25\hsize]{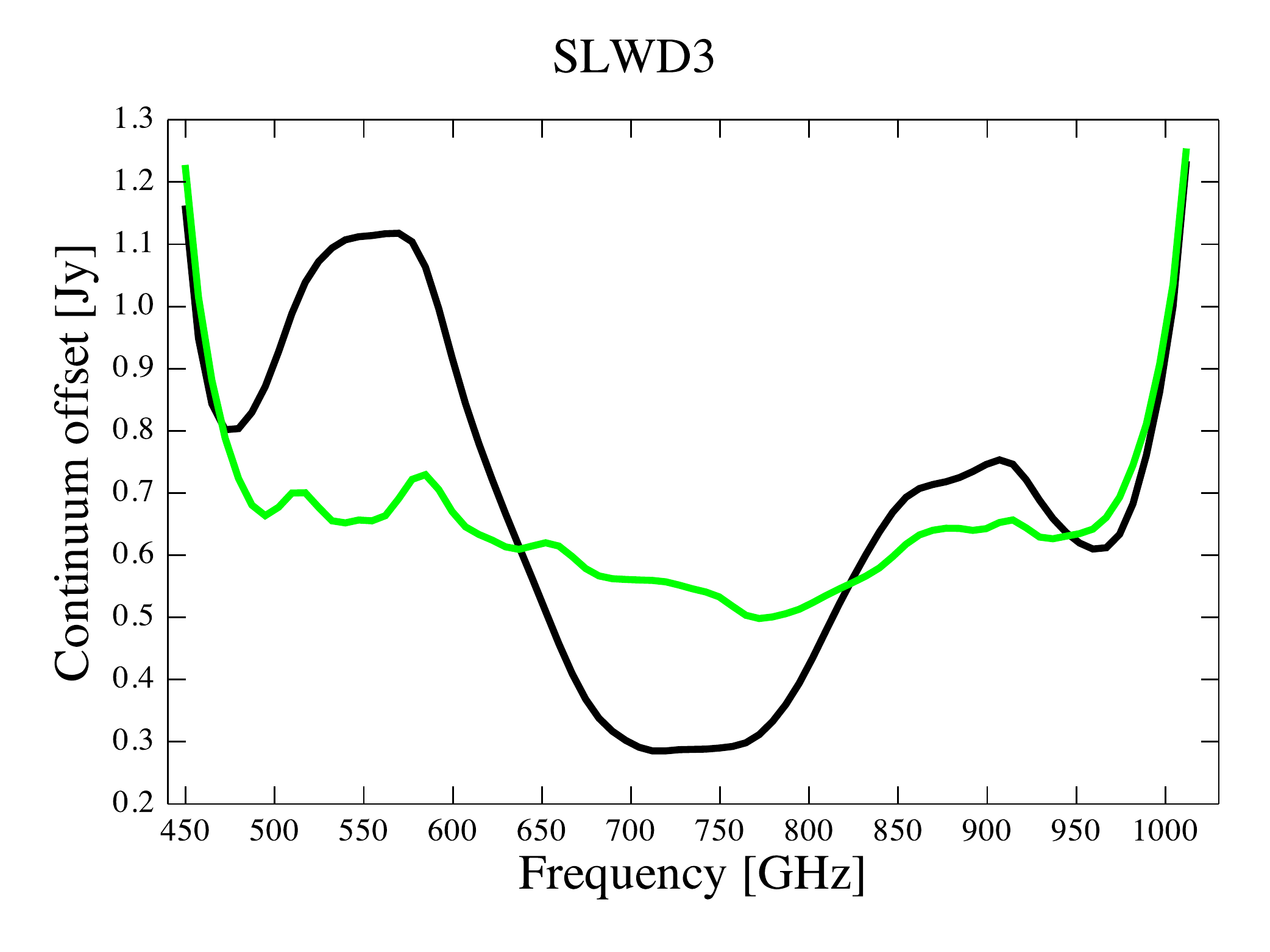}
\includegraphics[width=0.25\hsize]{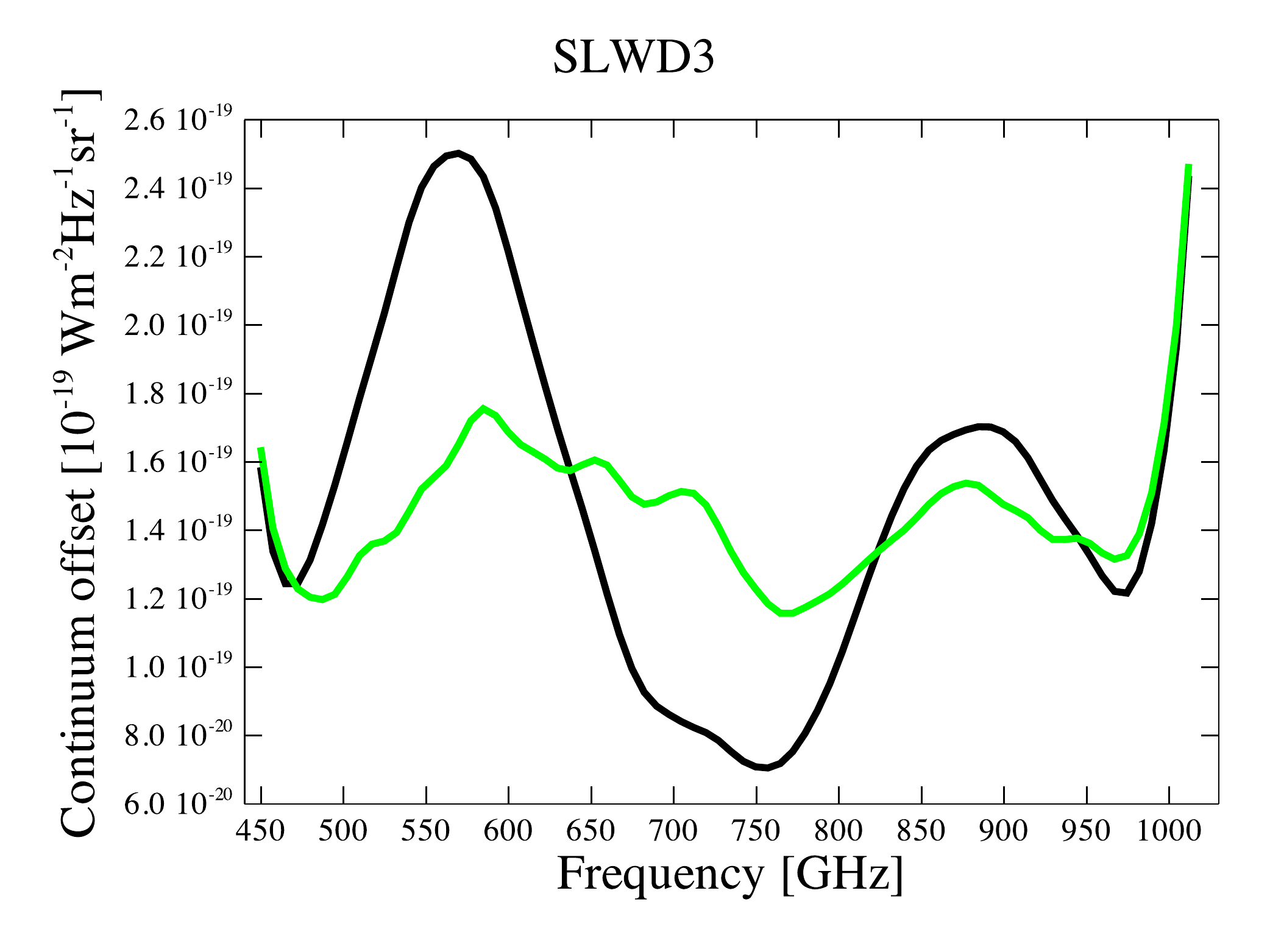}
\includegraphics[width=0.25\hsize]{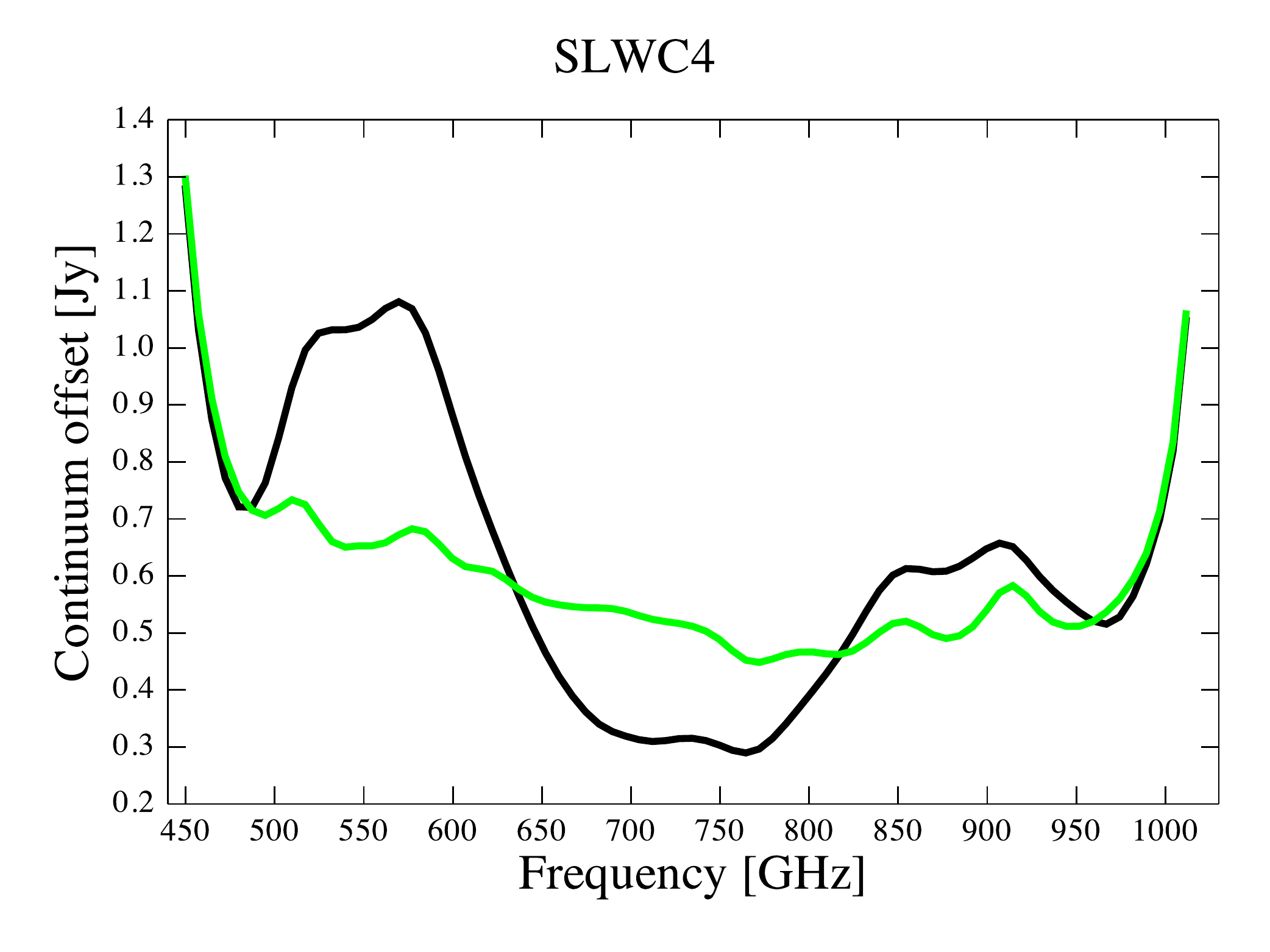}
\includegraphics[width=0.25\hsize]{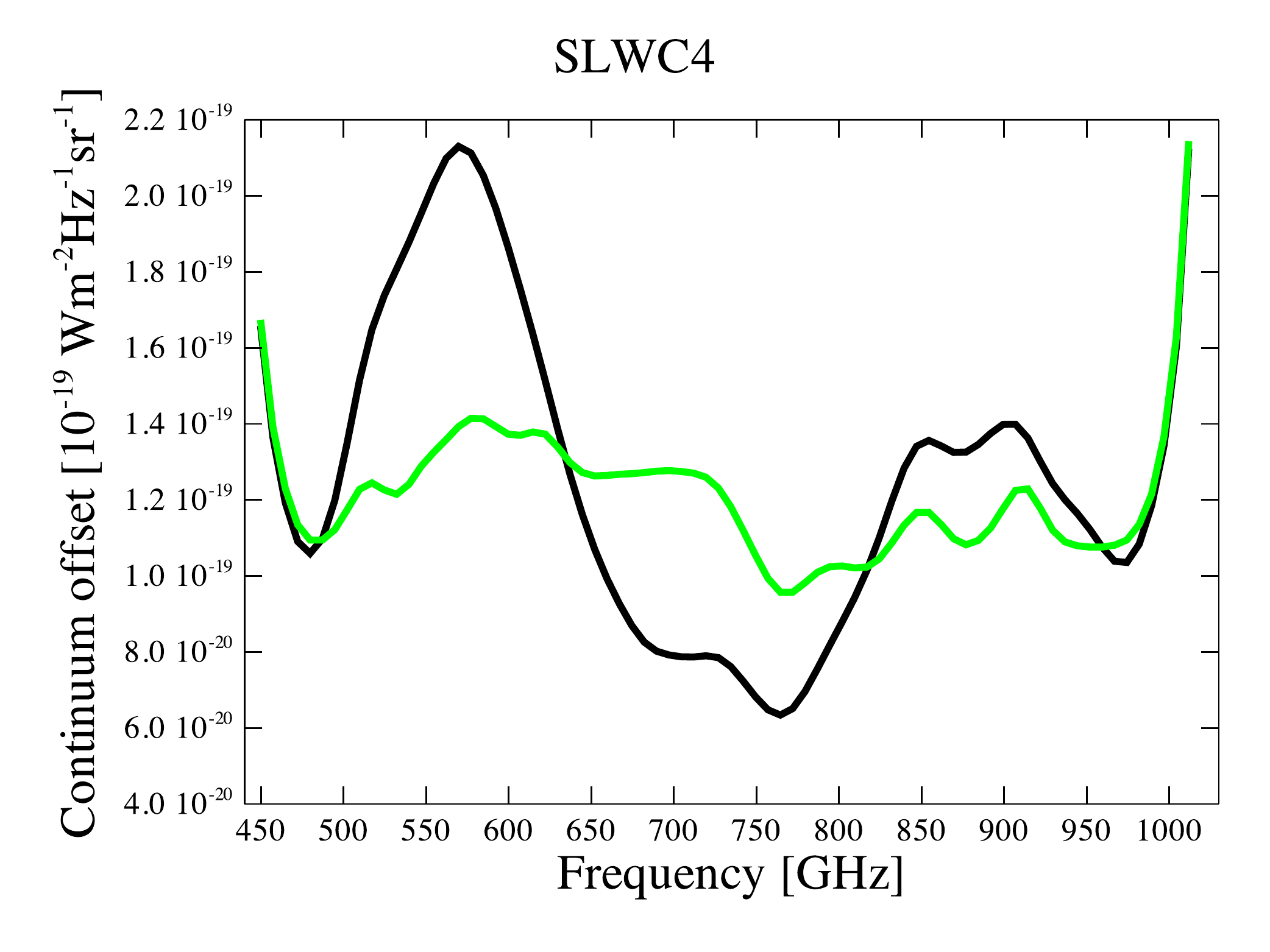}
}
\makebox{
\includegraphics[width=0.25\hsize]{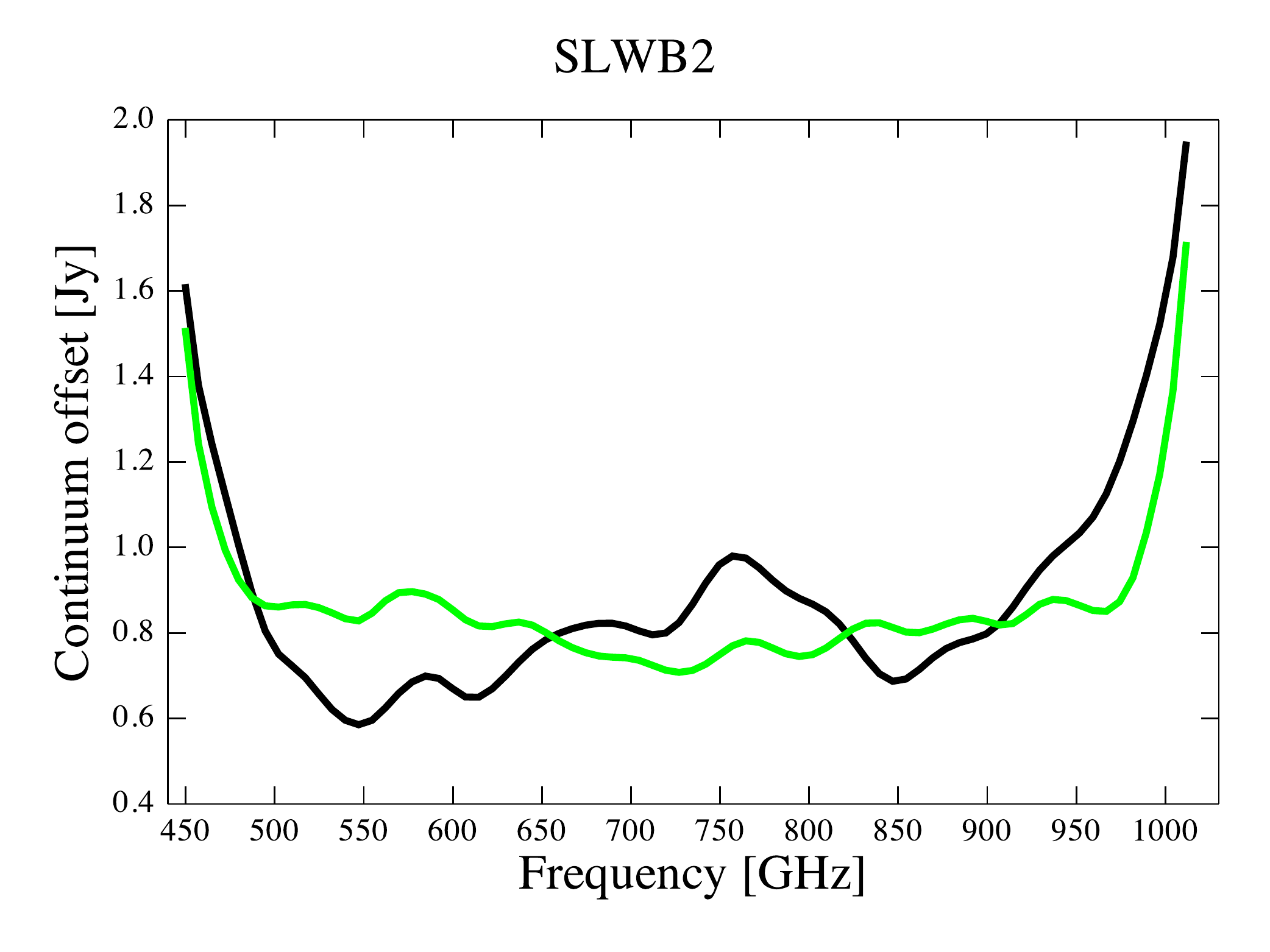}
\includegraphics[width=0.25\hsize]{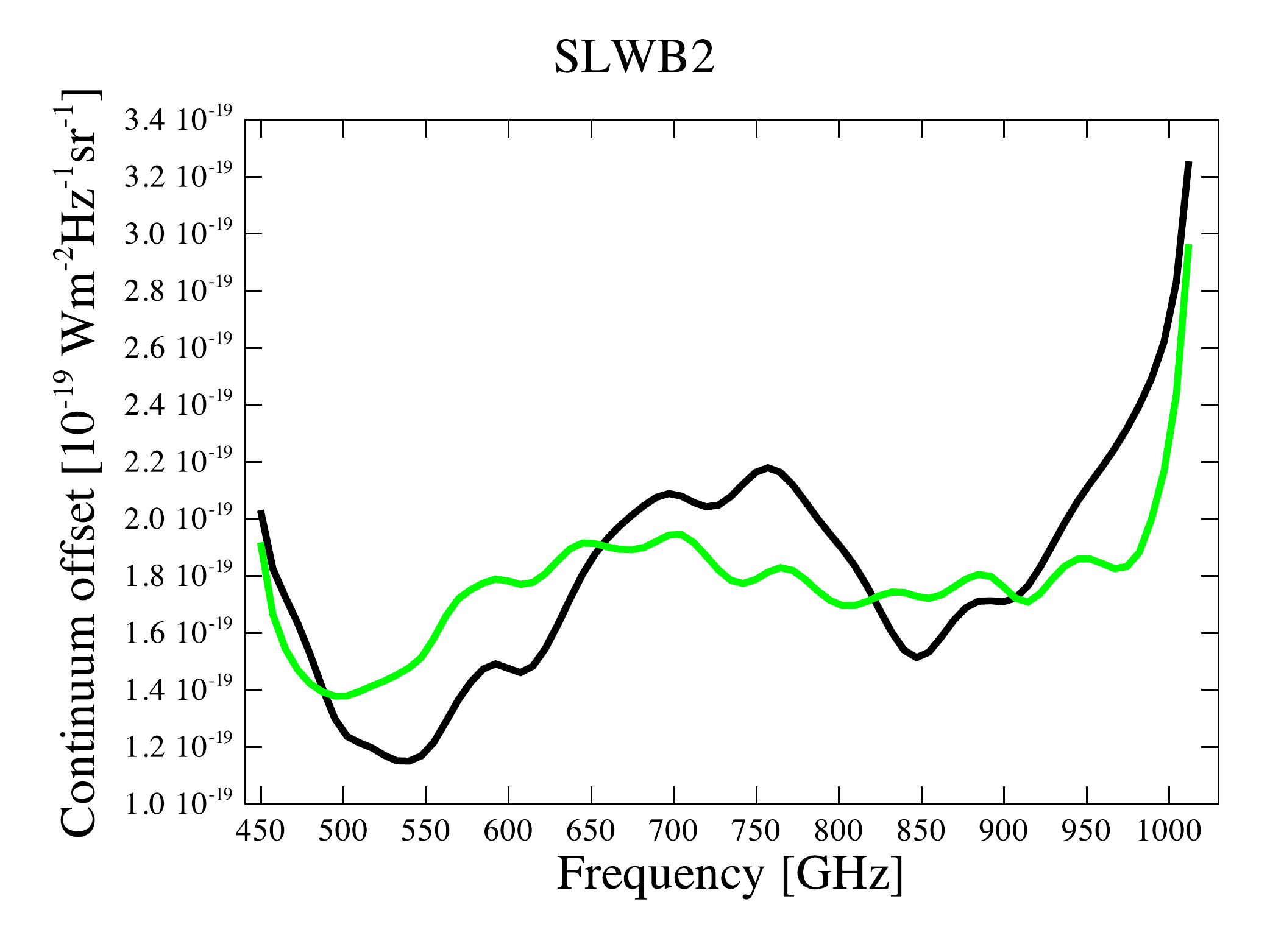}
\includegraphics[width=0.25\hsize]{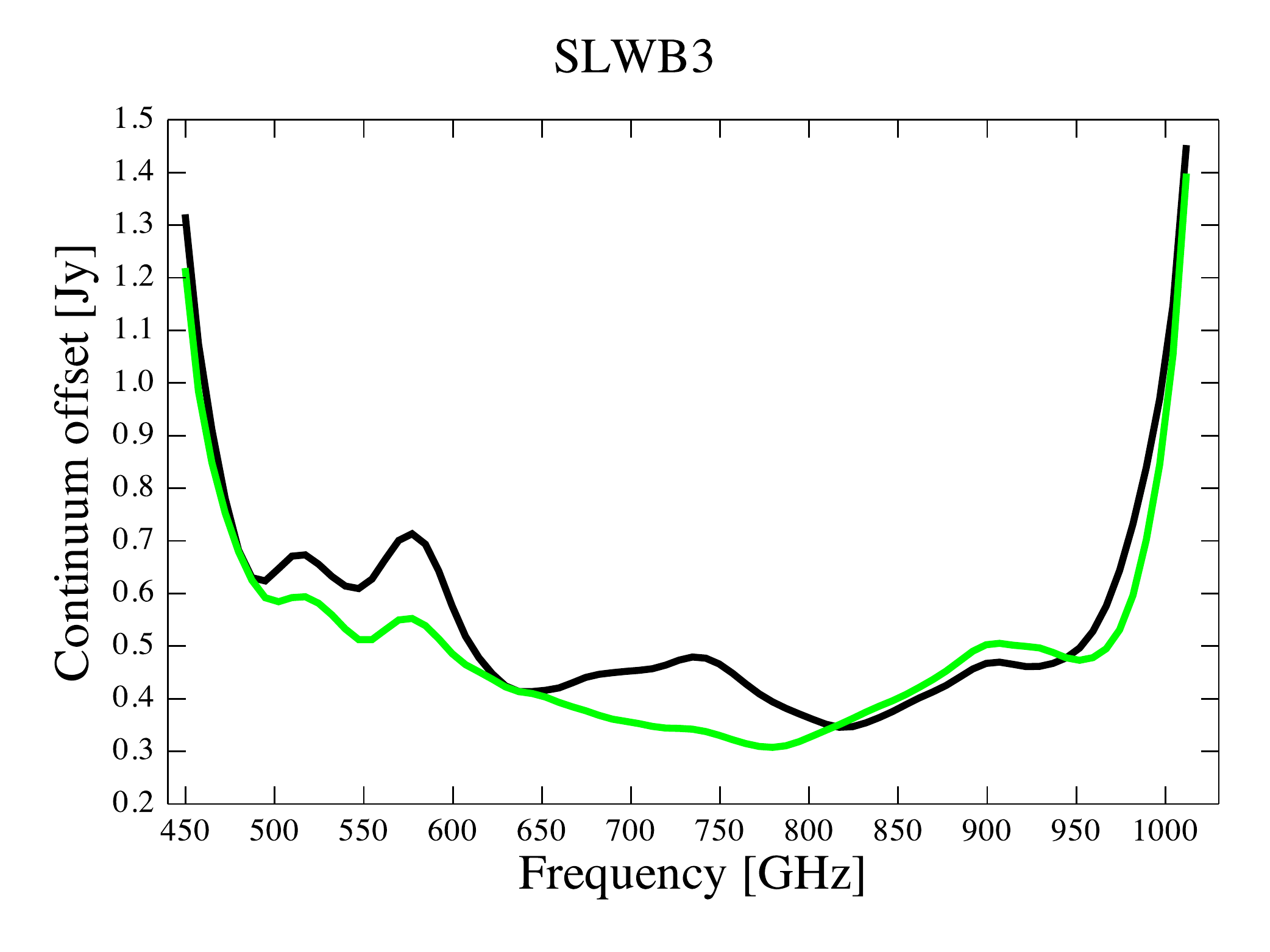}
\includegraphics[width=0.25\hsize]{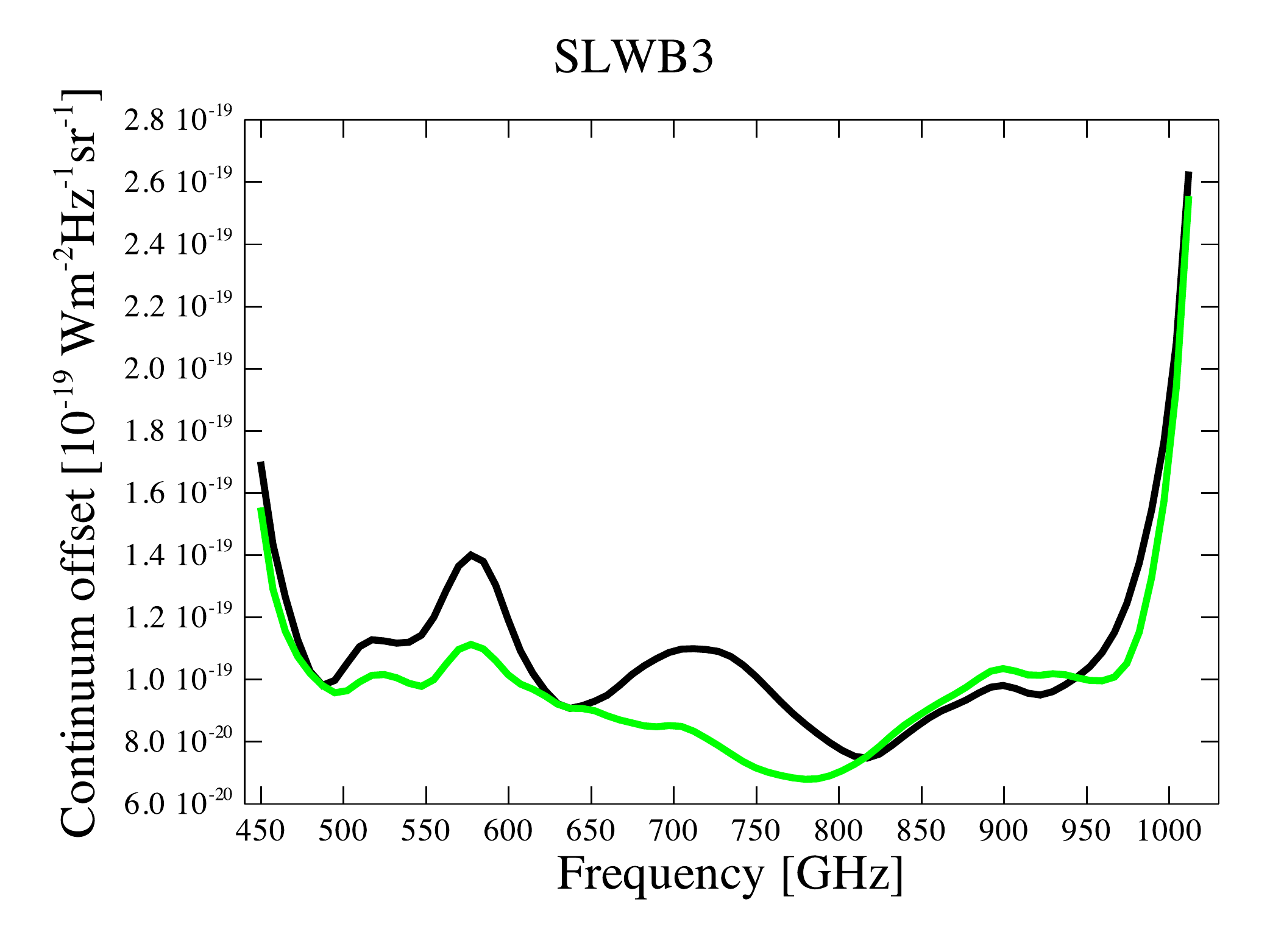}
}
\makebox{
\includegraphics[width=0.25\hsize]{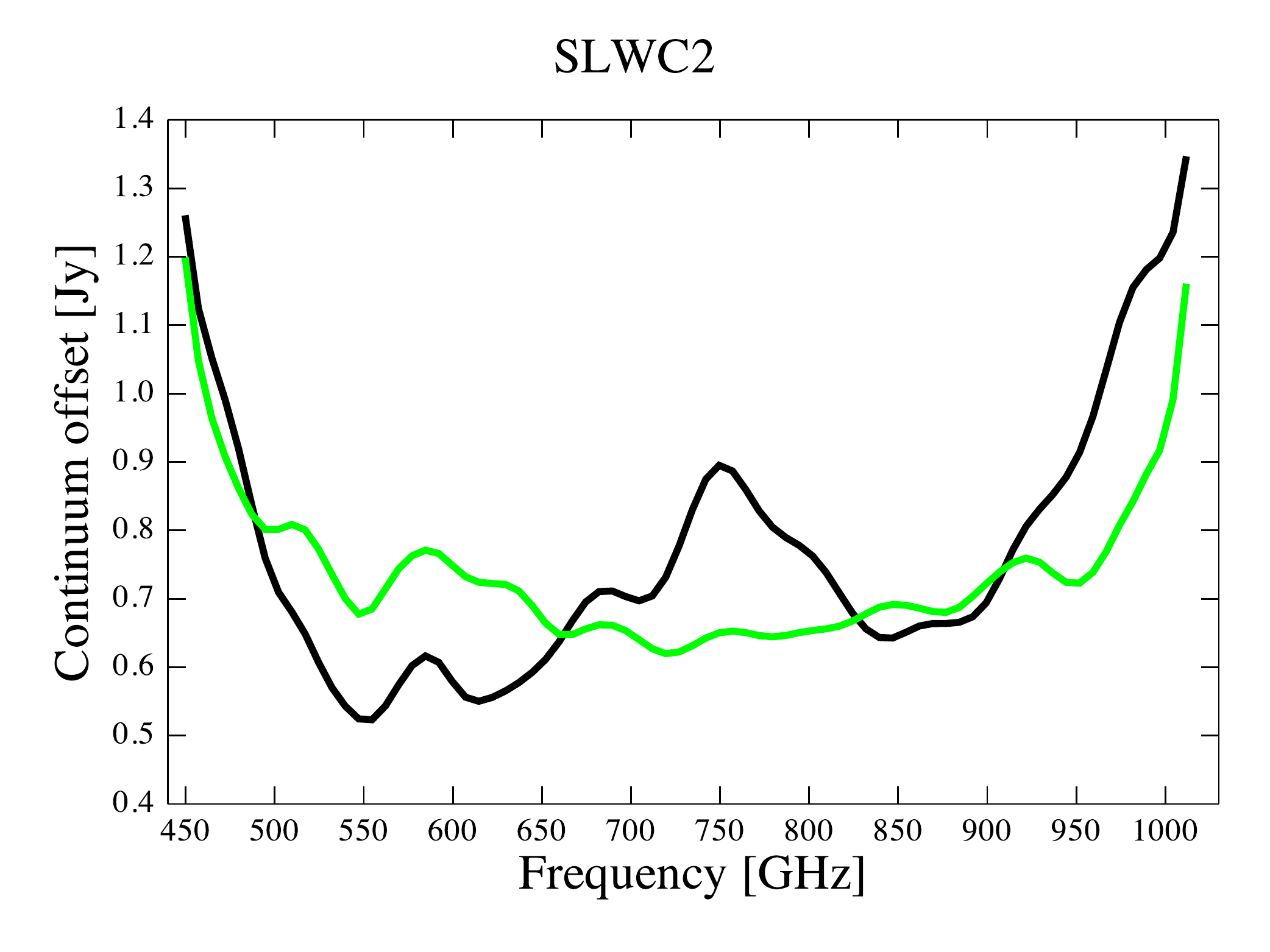}
\includegraphics[width=0.25\hsize]{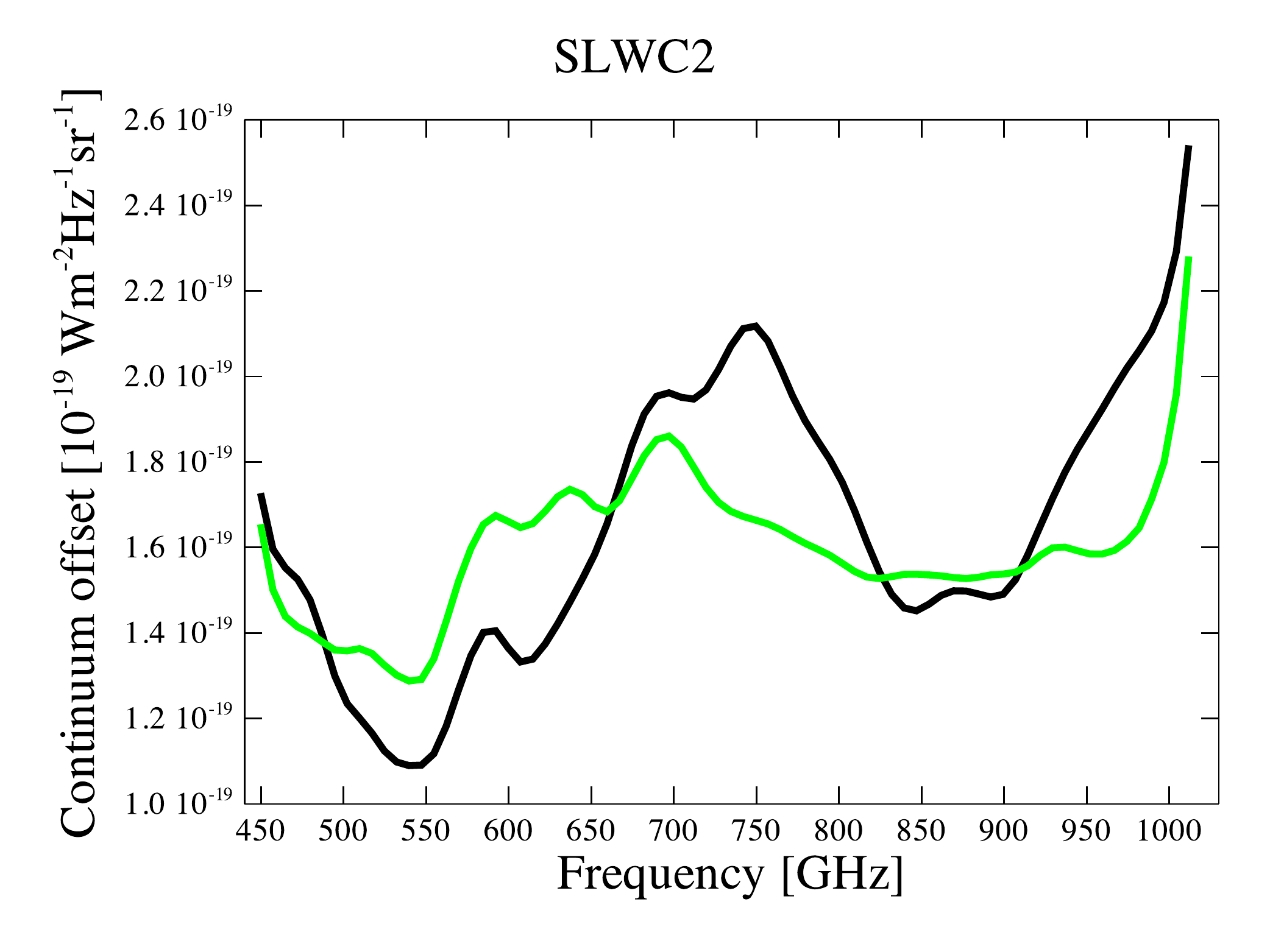}
}
\end{center}
\caption{LR continuum offset: for each detector, the offsets for point-source calibrated data (left) and extended-source calibrated data (right), before (black) and after (green) application of the LR correction are shown.}
\label{fig:LRoffs1}
\end{figure*}



\bsp	
\label{lastpage}

\begin{thebibliography}{99}
\bibitem[\protect\citeauthoryear{Ade, Hamilton \& Naylor}{1999}]{Ade99} Ade P., Hamilton P., Naylor D., 1999, Fourier Transform Spectroscopy: New Methods and Applications. OSA Technical Digest, Optical Society of America, paper FWE3
\bibitem[\protect\citeauthoryear{Dohlen et al.}{2000}]{Dohlen00}
    Dohlen, K., Origne, A., Pouliquen, D., \& Swinyard, B.~M.\, 2000, \procspie, 4013, 119 
\bibitem[\protect\citeauthoryear{Fulton et al.}{2010}]{Fulton10} Fulton, T.~R., Baluteau, J.-P., Bendo, G., et al.\ 2010, \procspie, 7731, 773134
\bibitem[\protect\citeauthoryear{Fulton et al.}{2014}]{Fulton14} Fulton, T., Hopwood, R., Baluteau, J.-P., et al.\ 2014, Experimental Astronomy, 37, 381
\bibitem[\protect\citeauthoryear{Fulton et al.}{2016}]{Fulton16} Fulton, T., Naylor, D. A., Polehampton, E. T., et al. \ 2016, \mnras, 458, 1977
\bibitem[\protect\citeauthoryear{Griffin et al.}{2010}]{Griffin10} Griffin, M.~J., Abergel, A., Abreu, A., et al.\ 2010, \aap, 518, L3
\bibitem[\protect\citeauthoryear{Hopwood et al.}{2014}]{Hopwood14} Hopwood, R., Fulton, T., Polehampton, E.~T., et al.\ 2014, Experimental Astronomy, 37, 195
\bibitem[\protect\citeauthoryear{Hopwood et al.}{2015}]{Hopwood15} Hopwood, R., Polehampton, E.~T., Valtchanov, I., et al.\ 2015, \mnras, 449, 2274
\bibitem[\protect\citeauthoryear{Ott}{2010}]{Ott10} Ott, S.\ 2010, Astronomical Data Analysis Software and Systems XIX, 434, 139
\bibitem[\protect\citeauthoryear{Pilbratt et al.}{2010}]{Pilbratt10} Pilbratt, G.~L., Riedinger, J.~R., Passvogel, T., et al., 2010, A\&A, 518, L1
\bibitem[\protect\citeauthoryear{SPIRE Handbook}{2016}]{SPIRE16} SPIRE Handbook 2016, HERSCHEL-HSC-DOC-0798, accessed from http://herschel.esac.esa.int/Documentation.shtml
\bibitem[\protect\citeauthoryear{Swinyard et al.}{2003}]{Swinyard03} Swinyard, B.~M., Dohlen, K., Ferand, D., et al.\ 2003, \procspie, 4850, 698 
\bibitem[\protect\citeauthoryear{Swinyard et al.}{2010}]{Swinyard10} Swinyard, B.~M., Ade, P., Baluteau, J.-P., et al.\ 2010, \aap, 518, L4 
\bibitem[\protect\citeauthoryear{Swinyard et al.}{2014}]{Swinyard14} Swinyard, B.~M., Polehampton, E.~T., Hopwood, R., et al.\ 2014, \mnras, 440, 3658 
\bibitem[\protect\citeauthoryear{Turner et al.}{2001}]{Turner01} Turner, A.~D., Bock, J.~J., Beeman, J.~W., et al.\ 2001, \ao, 40, 4921 
\end{thebibliography}
\end{document}